\documentclass[prb,twocolumn,superscriptaddress]{revtex4-2}
\usepackage[utf8]{inputenc}
\usepackage[T1]{fontenc}
\usepackage{graphicx}
\usepackage{amsmath}
\usepackage{amssymb}
\usepackage{bm}
\usepackage{hyperref}
\usepackage{pgf}
\usepackage{comment}
\usepackage{chngcntr}
\usepackage[normalem]{ulem}

\newcommand*{\parallelogramm}
{ \rlap{\rotatebox{-30}{\rule[.05ex]{.4pt}{.77em}}}
  \kern.04em
  \rlap{\kern.36em\raisebox{0.649519052835em}{\rule{.6em}{.4pt}}}
  \rule{.6em}{.4pt}\kern-.04em
  \rotatebox{-30}{\rule[.05ex]{.4pt}{.77em}}}

\begin{document}
\title{Charge disproportionation as a possible mechanism towards polar antiferromagnetic metal in molecular orbital crystal}
\author{Yang Shen}
\affiliation{State Key Laboratory of Quantum Functional Materials and School
of Physical Science and Technology, ShanghaiTech University, Shanghai 201210, China}
\author{Shuai Qu}
\affiliation{State Key Laboratory of Quantum Functional Materials and School
of Physical Science and Technology, ShanghaiTech University, Shanghai 201210, China}
\author{Gang Li}
\email{ligang@shanghaitech.edu.cn}
\affiliation{State Key Laboratory of Quantum Functional Materials and School
of Physical Science and Technology, ShanghaiTech University, Shanghai 201210, China}
\author{Pu Yu}
\affiliation{State Key Laboratory of Low-Dimensional Quantum Physics and Department of
Physics, Tsinghua University, Beijing 100084, China}
\author{Guang-Ming Zhang}
\email{zhanggm@shanghaitech.edu.cn}
\affiliation{State Key Laboratory of Quantum Functional Materials and School
of Physical Science and Technology, ShanghaiTech University, Shanghai 201210, China}
\affiliation{Department of Physics, Tsinghua University,
Beijing 100084, China}
\date{\today}

\begin{abstract}
Polar antiferromagnetic metals have recently garnered increasing interests due to their combined traits of both ferromagnets (with spin polarization) and antiferromagnets (absence of net magnetization) for spintronic applications. However, the inherently incompatible nature of antiferromagnet, metallicity and polarity pose a significant challenge in designing these meterials. Herein we propose that charge disproportionation can lead to this novel state in negative charge transfer gap regime in molecular orbital crystal.
As a proof of concept, this proposal is demonstrated by molecular orbital analyses of first-principles DFT+$U$ electronic band structure for representative Ruddlesden-Popper bilayer perovskite oxides Sr$_3$Co$_2$O$_7$, corroborated by Density Matrix Renormalization Group calculation. Due to the negative charge transfer nature of Co$^{4+}$ and imposed by strong interlayer coupling, localized molecular orbitals stemming from the hybridization of Co $d_{z^2}$ and $d_{xz/yz}$ orbitals through the apical oxygen $p$ orbitals are preferably emergent within each bilayer unit, which develop antiferromagnetic ordering by invoking Hubbard repulsion. Charge disproportionation driven by Hund's physics, makes an occupation imbalance with broken inversion symmetry in the remaining $d_{xy}$ and $d_{x^2-y^2}$ orbitals from distinct Co atoms within the bilayer unit, resulting in the polar metallicity. Meanwhile, this charge disproportionation scenario allows consequent conducting carriers to couple with interlayer local spins via Hund's coupling, giving rise to in-plane double-exchange ferromagnetism. Our molecular orbital formulation further provides a guide towards an effective Hamiltonian for modelling the unconventional synergy of metallicity, polarity and antiferromagnetism in Sr$_3$Co$_2$O$_7$, which may be a unified framework widely applicable to double-layer Ruddlesden-Popper perovskite oxides.
\end{abstract}

\maketitle
\textit{Introduction}-- Integrating multiple order parameters that are conventionally incompatible within a single-phase material, enables the emergence of exotic phenomena and enhanced functionalities. Recently, polar antiferromagnetic metals\cite{PhysRevB.102.104410,lei2018observation} stand out as a novel paradigm for investigating exotic electronic state emerging from the mutually exclusive coupling between polarity, metallicity, and antiferromagnetism. In particular, the breaking of joint parity-time-reversal ($\mathcal{PT}$) symmetry in these materials may give rise to the anomalous Hall effect with non-vanishing Berry curvature distribution, a phenomenon typically associated with ferromagnets and more profound in altermagnets\cite{PhysRevX.12.031042}. However, the prevalence of ferromagnetic order in metallic materials, coupled with the inherently incompatible nature of metallicity\cite{Song2023,PhysRevLett.78.1751} and polarity\cite{anderson1965symmetry,kim2016polar,bhowal2023polar}, pose a significant challenge in designing and discovering polar antiferromagnetic metals.

\begin{figure}[t]
\includegraphics[width=0.5\textwidth]{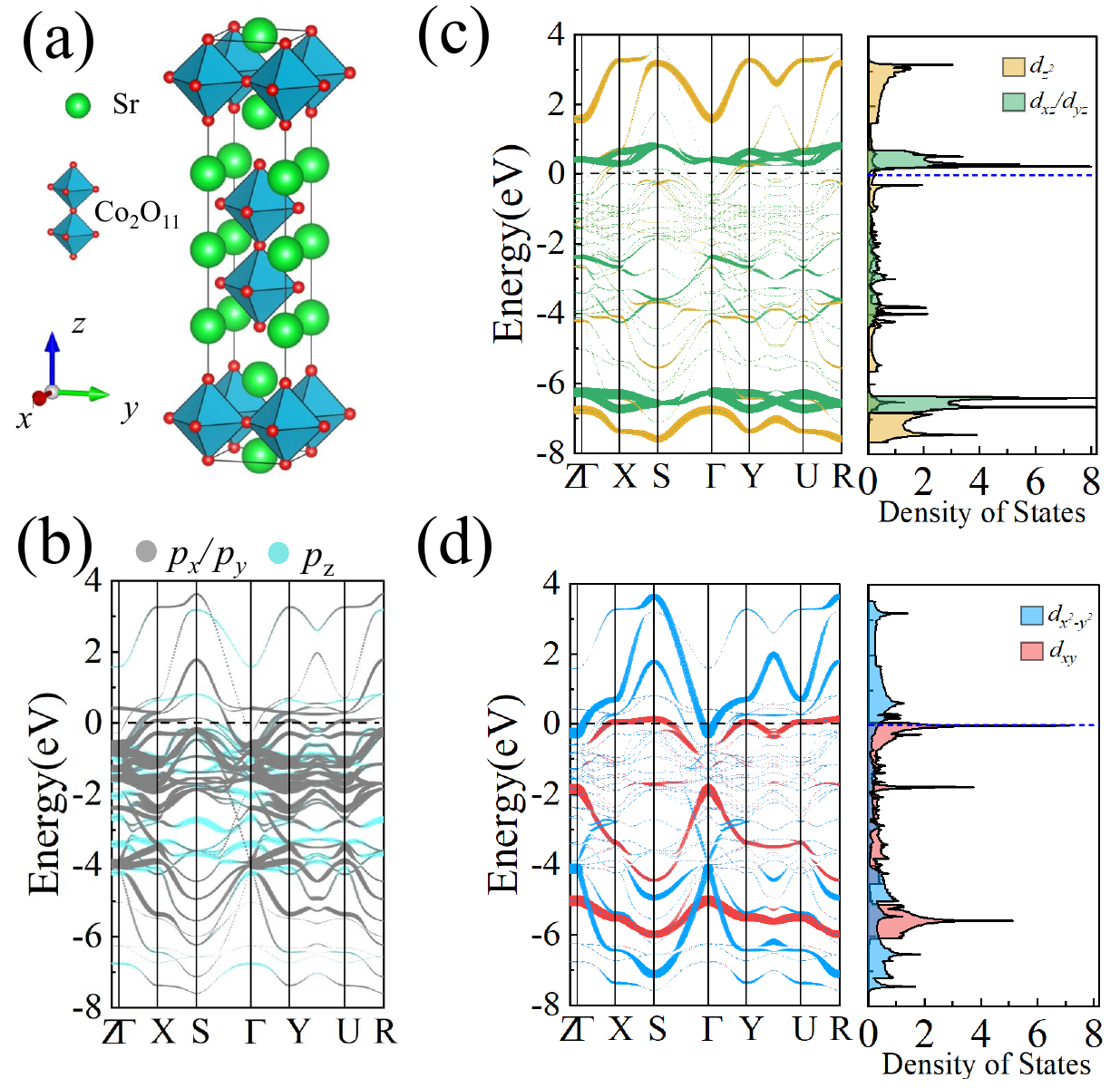}
\caption{(a) Schematic crystal structure of the conventional unit cell (I$4mm$, space group No. 107) and (b-d) the orbital-weight-projected DFT band structures along with their corresponding density of states, for the A-type antiferromagnetic ferroelectric phases of the Ruddlesden-Popper phase cobaltate Sr$_3$Co$_2$O$_7$. In the conventional unit cell, the local $x-$ or $y$-axis is along the in-plane Co-O bond directions while the local z-axis is along the out-of-plane Co-O bond directions. The characteristic adjacent Co-O octahedra linked by apical O anions (designated by Co$_2$O$_{11}$) are depicted as interconnected octahedra structures in (a). The electronic structures are grouped into out-of-plane and in-plane orbital species for (b) O $2p$ and (c-d) Co $3d$ orbitals, and these orbitals' weights are indicated by the size of the circles in the band structures. Notice that O $2p_x$ and $2p_y$, as well as Co $3d_{xz}$ and $3d_{yz}$ are symmetrically equivalent by $C_4$ rotation operation. The Fermi level, denoted as zero energy, is marked by the horizontal dashed line.
}
\label{band_structure}
\end{figure}

Double-layered Ruddlesden-Popper (RP) perovskite oxides A$_3$B$_2$O$_7$ have attracted significant attention due to their remarkable physical properties, including colossal magnetoresistance\cite{RevModPhys.73.583}, multiferroicity\cite{PhysRevLett.106.107204,doi:10.1126/science.1262118}, ferroelectricity\cite{smith2019infrared,lee2013exploiting}, metamagnetic transitions\cite{lei2018observation}, and superconductivity\cite{sun2023signatures,ko2024signatures,wang2024bulk,zhou2024ambient}. Among them, a novel polar antiferromagnetic metallic state has been very recently identified in Sr$_3$Co$_2$O$_7$\cite{Yu}, where every two adjacent Co-O octahedrons is interconnected by apical oxygens with AB stacking sequence along the $z$ direction (see Fig.\ref{band_structure}(a)). Different off-centric displacements of Co ions (5.4 pm and -0.3 pm, respectively) in adjacent oxygen octahedrons break the inversion symmetry and introduce vertical polar axis\cite{Yu}, which interplays with A-type antiferromagnetic order and metallity to break the joint $\mathcal{PT}$ symmetry, leading to the pronounced anomalous Hall effect in experiments and identifying the material as a promising altermagnet candidate. Due to the quantum confinement of bilayer structure, Co $3d$ orbitals, which interact with the apical oxygen, tend to exhibit large interlayer coupling, thereby facilitating the formation of interlayer molecular orbitals\cite{shen2023effective,PhysRevB.108.L201108}. Given that, in the isostructural polar antiferromagnetic metal Ca$_3$Ru$_2$O$_7$\cite{sokolov2019metamagnetic,lei2018observation,doi:10.1073/pnas.2003671117,peng2024flexoelectric,gui2022improper}, inversion symmetry is lifted by octahedral rotations with emergent polarization inside the plane, it is intriguing to ask why polarity in Sr$_3$Co$_2$O$_7$ arises from the out-of-plane off-centric displacement of Co ions rather than from octahedral rotations, and how it interplays with the antiferromagntism and metallicity.

In this work, we propose that the unconventional synergy of metallicity, polarity and antiferromagnetism in Sr$_3$Co$_2$O$_7$ originates from unique bilayer architecture with pronounced orbital hybridization and Hund's physics, which exhibits orbital selective behaviors from localized and delocalized electrons. Specifically, the emergent interlayer antiferromagnetic ordering can be attributed to the formation of localized molecular orbitals arising from quantum-confinement-induced hybridization between Co $d_{z^2}$ and $d_{xz/yz}$ orbitals, mediated by bridging apical oxygen $p$ orbitals. Crucially, charge disproportionation, governed by Hund’s physics, induces an occupancy imbalance between the remaining $d_{xy}$ and $d_{x^2-y^2}$ orbitals across distinct Co ions within the bilayer, thereby establishing metallic charge polarity. This subtle charge redistribution further enables these conducting carriers to interact with localized interlayer spins, generating in-plane ferromagnetic ordering via double-exchange mechanism.

\textit{Negative charge transfer gap}-- To illuminate this idea, we develop a microscopic theory to elucidate the formation of localized molecular orbitals with charge disproportionation, and their interplays with itinerant atomic orbitals. We start our discussions on the emergent local molecular orbitals by first understanding the valence configuration of Co$^{4+}$. The calculated DFT+$U$ band structures along with their corresponding density of states are presented in Fig. \ref{band_structure}(b-d) (see methods in section A in Supplementary Materials). Herein we adopt $U_{eff}$=4 within Dudarev's approach, which is close to the estimated values $U_{cRPA}=4.4 eV, U_{cRPA}’ =2.7 eV$ and $J_{H,cRPA}=0.85 eV$ from constrained random-phase approximation. It is clear that strong interlayer coupling leads to significant energy splitting of the $d_{z^2}$ and $d_{xz}/d_{yz}$ orbitals in Sr$_3$Co$_2$O$_7$, as shown in Fig. \ref{band_structure}(c). The $d_{z^2}$ band lies below the $d_{xz}/d_{yz}$ bands beneath the Fermi level, which is opposite to the typical octahedral crystal field splitting. This inversion indicates a significant $p-d$ orbital hybridization\cite{khomskii2014transition}, marking a departure from conventional octahedral field behavior. The O 2$p$ orbitals in Fig.\ref{band_structure}(b) have higher energy than $d_{z^2}$ and $d_{xz}/d_{yz}$ orbitals, which indicates a negative charge transfer gap in the Zaanen-Sawatzky-Allen classification scheme\cite{zaanen1985band}. Moreover, the $d_{x^2-y^2}$ and $d_{xy}$ orbitals in Fig.\ref{band_structure}(d) intersect with fermi level, resulting in slightly electron doping in $d_{x^2-y^2}$ orbital and hole doping in $d_{xy}$ orbital. The charge transfer from $d_{xy}$ to $d_{x^2-y^2}$ orbitals may be mediated by oxygen 2$p$ orbitals, as is evident from the significant energy overlap among these orbitals near the fermi level.

\begin{figure}[t]
\includegraphics[width=0.5\textwidth]{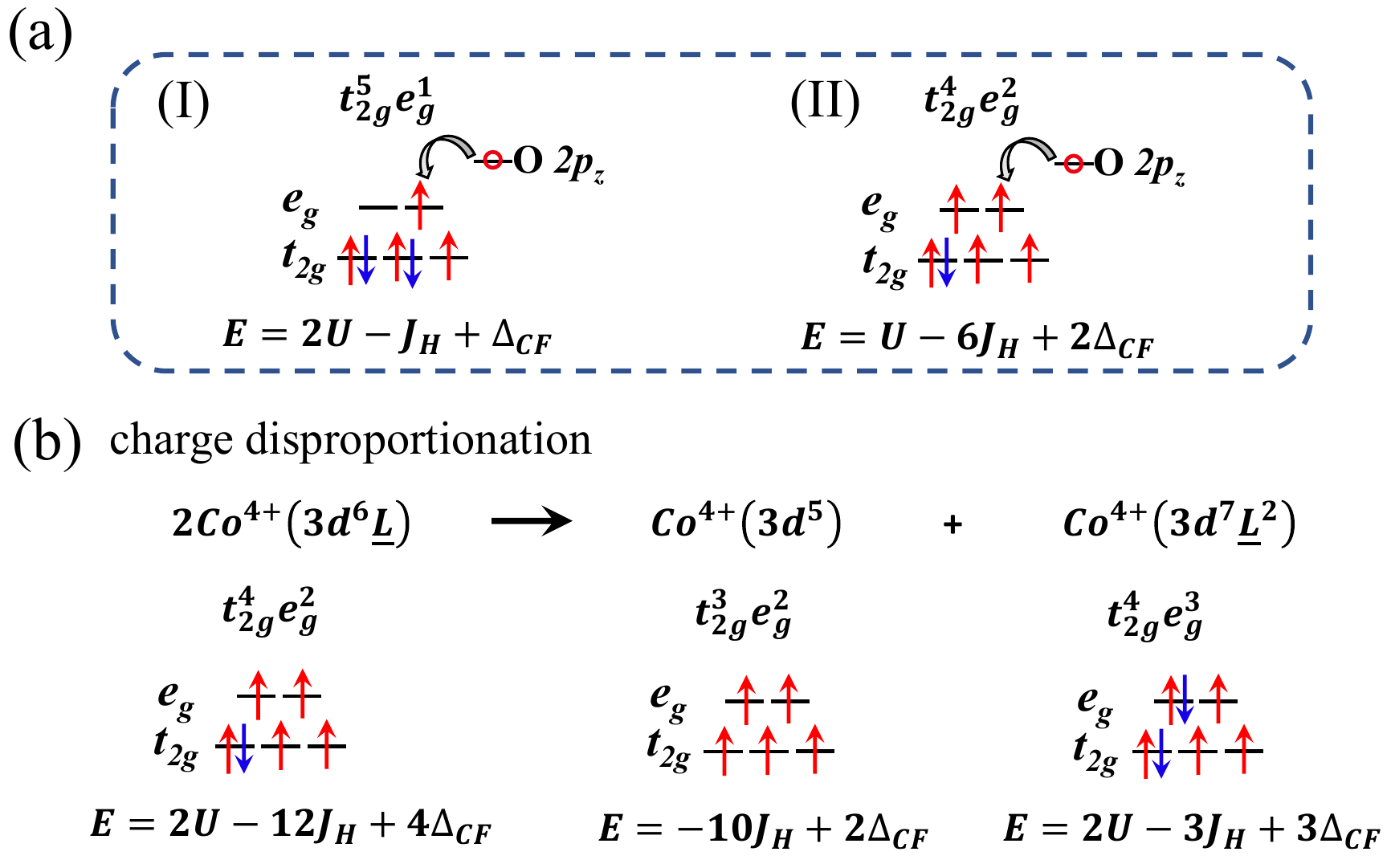}
\caption{Schematic electronic level diagram of 3$d^6 \underline L$ for (a) $t_{2g}^5e_g^1$ and $t_{2g}^4e_g^2$ electron configurations, and (b) charge disproportionation. The inset in (a) demonstrates the formation of Co 3$d^6 \underline L$ intermediate state by leaving the ligand hole (in red circle) in the oxygen 2$p_z$ orbital (coined as self-doping effect), since Co$^{4+}$ is in the negative charge transfer gap regime and interlayer quantum confinement. Due to the energy gain of Hund's coupling $J_H$, the Co$^{4+}$ (3$d^6 \underline L$ dimer) in the adjacent bilayer Co-O octahedra may exhibit spontaneous charge disproportionation, $i.e.$, 2Co$^{4+}$ ($3d^6\underline L,t_{2g}^4e_g^2$) $\xrightarrow{}$ Co$^{4+}$ ($3d^5, t_{2g}^3e_g^2$) + Co$^{4+}$ ($3d^7\underline L^2,t_{2g}^4e_g^3$), leading to the emergence of the polar metal with different cobalt atoms' off-centric displacive amplitudes and magnetic moments. }
\label{charge_dispro}
\end{figure}

Due to the negative charge transfer nature, 3$d^6 \underline L$ is a more favorable valence configuration than 3$d^{5}$\cite{PhysRevX.10.021030}.
As depicted in Fig.~\ref{charge_dispro}(a), this is achieved by preferentially donating an electron to the unoccupied $d_{z^2}$ orbital from the oxygen doubly occupied 2$p_z$ orbital which leaves a ligand hole in it in the case of large interlayer coupling in Sr$_3$Co$_2$O$_7$. This process is known as self-doping effect and is prevalent in nickelate superconductors\cite{PhysRevLett.112.106404,zhang2020self,nomura2022superconductivity}.

Within a simplified ionic picture, we note that there are two possible candidates configurations for nominal Co$^{4+}$ (3$d^6 \underline L$). One is $t_{2g}^5e_g^1$ configuration (with both doubly occupied $d_{xy}$ and $d_{xz}$ (or $d_{yz}$) orbitals after considering the $p$-$d$ hybridization, discuss later), and the other one is $t_{2g}^4e_g^2$ configuration (with doubly occupied $d_{xy}$ orbitals). The corresponding energy levels are shown in Fig. \ref{charge_dispro}(a). The doubly orbital degeneracy of the former configuration usually indicates in-plane Jahn-Teller lattice distortion or charge disproportionation instabilities in $d_{xz}$ and $d_{yz}$ orbitals, which yet are heavily suppressed by the epitaxial stress in thin films\cite{Yu}. The relevant energies can be estimated from the following atomic Hamiltonian:
\begin{eqnarray}
\mathcal{H}&=&U \sum_{ \alpha \sigma \neq \beta \sigma^{\prime}} n_{\alpha \sigma} n_{ \beta \sigma^{\prime}} -J_H \sum_{\alpha \neq \beta} \vec{S}_{\alpha} \vec{S}_{ \beta} \notag\\
&& + \Delta_{CF} \sum_{\alpha,\sigma} c_{ \alpha \sigma}^{\dagger} c_{\alpha \sigma}
\end{eqnarray}
Here $U$ represents the on-site Hubbard interactions, $J_H$ is the Hund's coupling and $\Delta_{CF}$ (=$\epsilon_{e_{g}}-\epsilon_{t_{2g}}$)  denotes octahedral crytal field splitting energy for 3$d$ orbitals. $\alpha$ and $\sigma$ represent $3d$ orbitals and spin indices, respectively. $\vec{S}$ is the $S=\frac{1}{2}$ operator for 3$d$ electrons. Compared to the $t_{2g}^5e_g^1$ configuration, the $t_{2g}^4e_g^2$ configuration has an atomic energy difference of $-U-5J_H+\Delta_{CF}$, $i.e.$, there is an additional energy loss of $\Delta_{CF}$ and an energy gain of $U$ and $J_H$. The final electronic configuration usually depends on the relative magnitudes of  $J_H$, $\Delta_{CF}$ and $U$.

\textit{Molecular orbital with charge disproportionation}--
With the local Co$^{4+} (3d^6\underline L)$ valence configuration, we now proceed to understand the Co-O-Co hybridization within each bilayer unit. Due to the interlayer coupling, $d_{yz}$, $d_{xz}$ and $d_{z^2}$ orbitals prefer to form molecular orbitals through the apical oxygen $p$ orbitals. As a consequence, the singly occupied $d_{yz}$ or $d_{xz}$ orbital has particularly large hybridization energy gain. As mentioned above, the energy splitting in Fig. \ref{band_structure}(c,d) that due to the $p-d$ hybridization is opposite to the regular point-charge octahedral crystal field, where $e_g$ levels lie below $t_{2g}$ multiplets underneath the fermi level. In this case, $\Delta_{CF}$ is significantly reduced, making high spin $t_{2g}^4e_g^2$ configuration with doubly occupied $d_{xy}$ orbitals more likely to be the ground state of the isolated Co-Co dimer in adjacent octahedra.

\begin{figure}[t]
\includegraphics[width=0.5\textwidth]{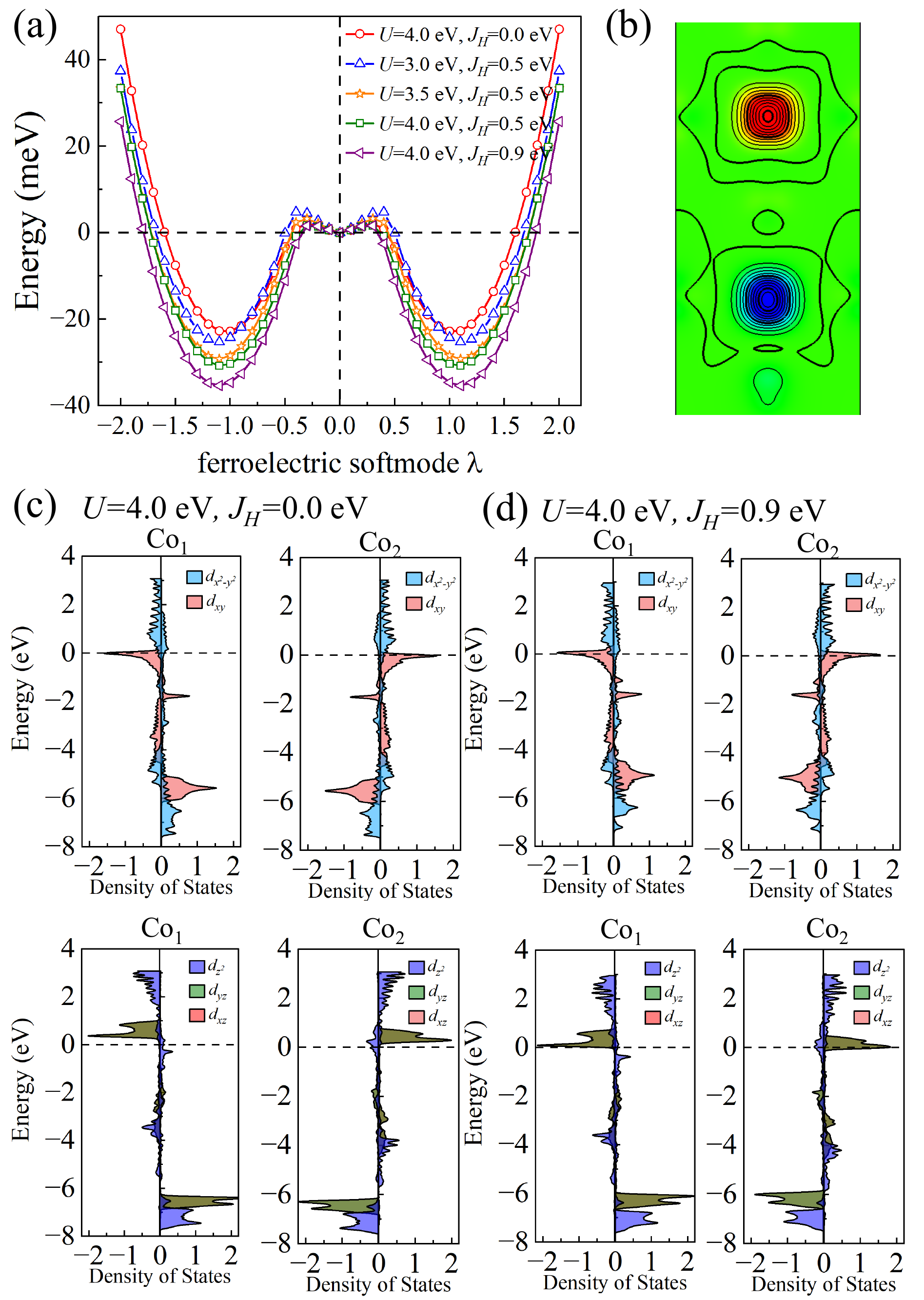}
\caption{The interplay between polar lattice distortion and charge disproportionation tuned by Hund's coupling.
	(a) Triplet-well potential profiles of Sr$_{3}$Co$_{2}$O$_{7}$, characterized by normalized displacive mode $\lambda$, wherein $\lambda = +1$ and $-1$ correspond to the referenced polarization ``up'' and ``down'' states for $U = 4.0\ \text{eV}$, $J_H = 0.0\ \text{eV}$.
	(b) Asymmetric spin density distribution with polarity of Co dimers from first-principles calculations. The spin majority and minority are colored in red and blue, respectively.
	(c, d) Orbital projected density of states for two inequivalent Co atoms at two representative Hund's coupling.}
\label{polar_mode}
\end{figure}

However, as shown in Fig.\ref{charge_dispro}(b), by developing a charge disproportionation, $i.e.$, 2Co$^{4+}$ ($3d^6\underline L,t_{2g}^4e_g^2$) $\xrightarrow{}$ Co$^{4+}$ ($3d^5, t_{2g}^3e_g^2$) + Co$^{4+}$ ($3d^7\underline L^2,t_{2g}^4e_g^3$), the Co-Co dimer system can have extra energy gain from Hund's coupling without energy loss of $U$. This process corresponds to one formal electron transfer from the $d_{xy}$ orbital of one Co ion to the $d_{x^2-y^2}$ orbital of another as shown in Fig.\ref{band_structure}(d) and Fig.\ref{polar_mode}(c,d). A more rigorous Density Matrix Renormalization Group study of Kanamori-type model on the simplified Co-O-Co cluster does find charge-disproportionate 3$d^6\underline L$ valence state is well established in the negative charge transfer gap regime with model parameters related to Sr$_3$Co$_2$O$_7$ (see section B in Supplementary Materials). As a consequence, these two charge-disproportionate Co$^{4+}$ ions tend to favor different chemical environments within each bilayer, e.g. displaying different off-centric displacements with charge polarity in Sr$_3$Co$_2$O$_7$. The charge-disproportionation driven polarity ultimately depends on the competition between the energy gain from Hund's coupling $J_H$ and elastic energy cost. Although in mixed-valence metallic systems electron screening tends to weaken dipole-dipole interactions and thereby reduce charge-disproportionation driven polarity, in confined layered systems such as Sr$_3$Co$_2$O$_7$ charge transport is restricted to the in-plane direction. As a result, the out-of-plane polarity is naturally protected by the out-of-plane charge gap and thus remains robust. Thus, there may be a crossover region where the charge disproportionation with polarity is stable with respect to high spin $t_{2g}^4e_g^2$ configuration. We can see that happens in Sr$_3$Co$_2$O$_7$ with Co$^{4+}$ (3$d^6 \underline L$) dimer in the adjacent bilayer Co-O octahedra.

To demonstrate this more explicitly, we use a $\lambda$-method (as described in the section E of Supplemental Materials) to describe polar instability characteristic of spontaneous symmetry-breaking soft phonon modes by tuning Hund's coupling strengths within the Lichtenstein\cite{PhysRevB.52.R5467} approach as implemented in VASP, which essentially captures the energy competition between Hund's coupling and elastic energy, and unveiling complicated interplay between polar lattice distortion, magnetic ordering, and charge ordering. The results are shown in Fig.\ref{polar_mode}.We first vary Hubbard $U$ parameter from 3, 3.5, to 4 eV at fixed Hund's coupling $J_H$=0.5 eV as shown in Fig.\ref{polar_mode}(a), the polarity (as explicitly shown in the asymmetric antiferromagnetic spin density distribution in Fig.\ref{polar_mode}(b)) remains almost unchanged, as evidenced by the constant energy minimum position at $\lambda$ = ±1.1 and only a slightly deepening triplet-well potential. In contrast, when Hund's coupling $J_H$ parameter is varied from 0, 0.5, to 1 eV with Hubbard $U$ fixed at 4 eV, the polarity is markedly enhanced, evidenced by a shift of  $\lambda$ from $\pm$1 to $\pm$1.1 and a significantly deepened triplet-well potential (see Fig.\ref{polar_mode}(b)). The corresponding orbital projected density of states on two different Co atoms are also shown for different $J_H$ values at fixed $U=4$ eV in Fig.\ref{polar_mode}(c,d). Take $J_H$=0.0 and 0.9 eV as examples. The $d_{xy}$ orbital is nearly fully occupied, while the remaining 3$d$ orbitals are close to half‑filled, approaching a high‑spin $t_{2g}^4 e_g^2$ configuration. However, the projected density of states on $d_{x^2-y^2}$ and $d_{xy}$ orbitals are not equivalent near the Fermi level for two Co atoms and become further different with the increase of $J_H$. A similar asymmetry even appears for the  $d_{z^2}$, $d_{xz}$ and $d_{yz}$ at larger $J_H$. The trend is further confirmed by roughly counting the 3$d$ orbital occupation numbers numerically obtained using trapezoidal integration rule, which show a growing difference in total 3$d$ electron count as $J_H$ rises (see Tab.\ref{table1}). The electron redistribution takes place in the manner that the spin polarization of $t_{2g}$ electrons increases and that of $e_g$ electrons decreases at Co$_1$, whereas both $t_{2g}$ and $e_g$ electron spin polarization decrease at Co$_2$, to maximizes the Hund’s energy gain in the Co dimers. The change tuned by Hund’s coupling is consistent with the proposed charge disproportionation mechanism, where the charge-disproportionate Co-Co dimer system can have extra energy gain from Hund’s coupling.  Therefore, a large Hund's coupling $J_H$ can overcome the elastic energy cost to make the charge disproportionation happen in Sr$_3$Co$_2$O$_7$.

\begin{table}[htbp]
    \centering
    \caption{The change of electron occupations of relevant 3$d$ orbitals of two inequivalent Co atoms, with Hund's coupling $J_H$ from first-principles calculation. The electron numbers of total 3$d$ electron, spin polarized $t_{2g}$, $e_g$ and total 3$d$ electrons of Co atoms are listed.}
     \label{table1}
    \includegraphics[width=0.5\textwidth]{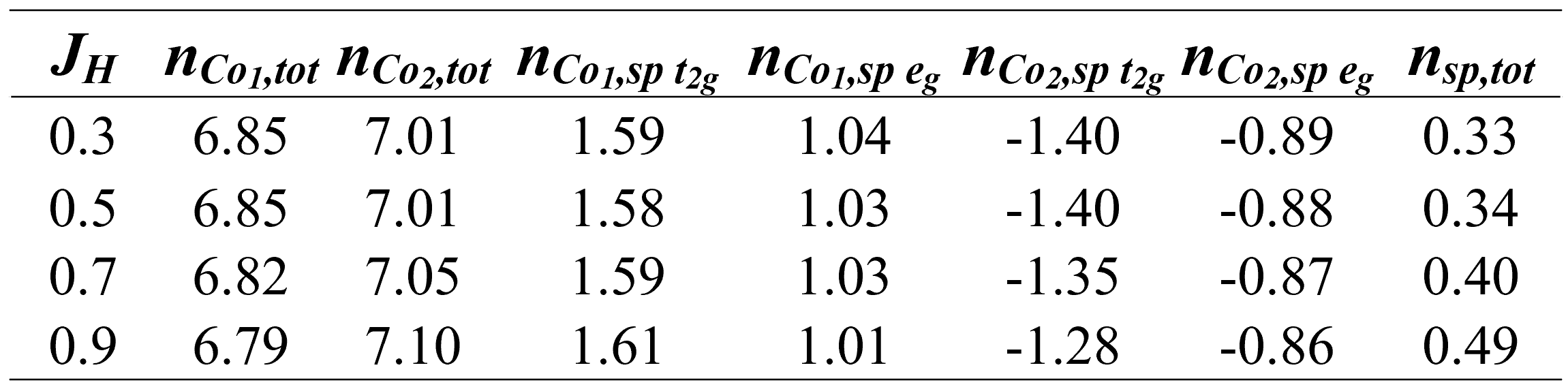}
\end{table}

\begin{figure}[t]
\includegraphics[width=0.5\textwidth]{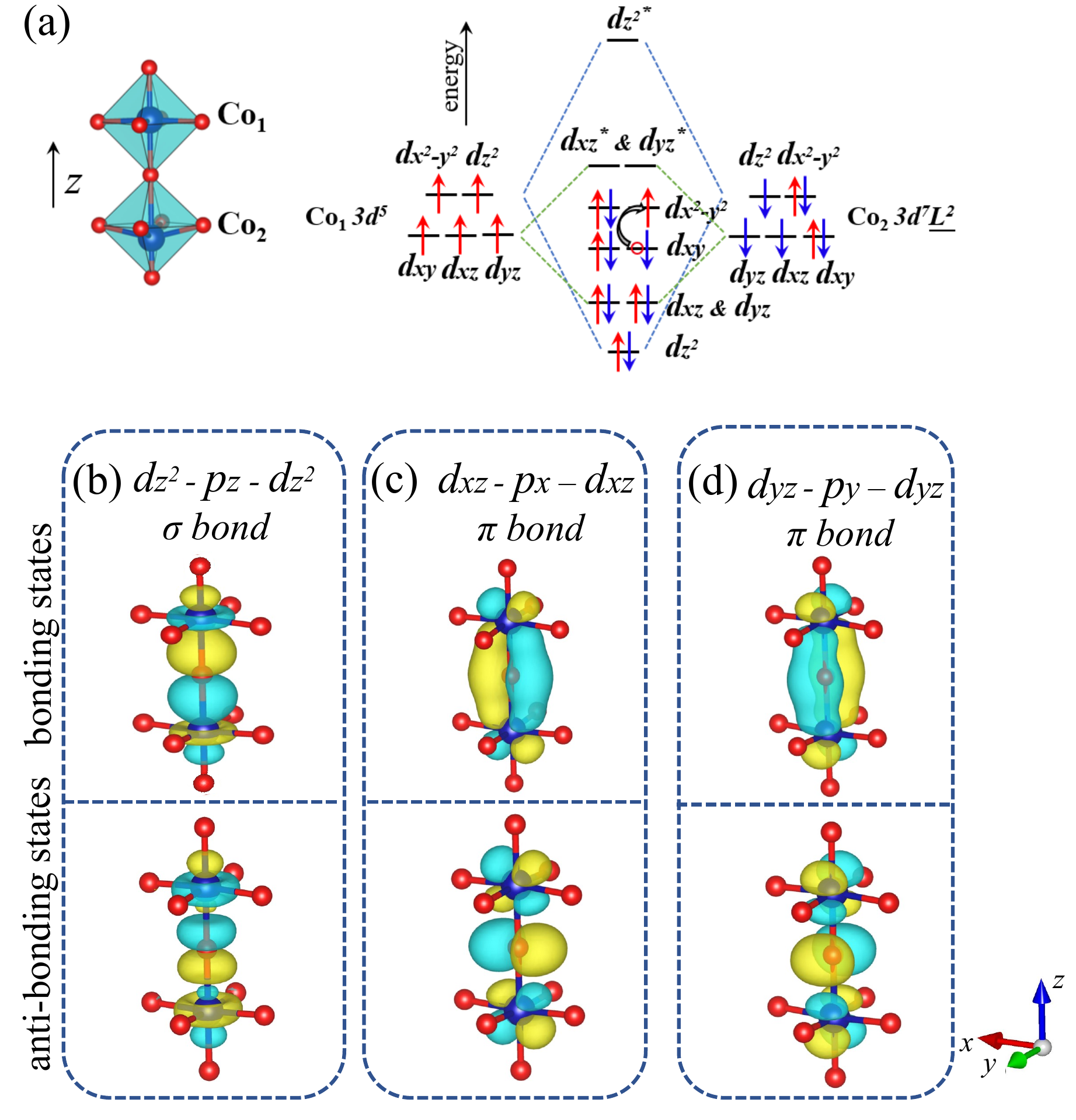}
\caption{(a) Energy diagram of the adjacent bilayer CoO octahedra in cobaltate Sr$_3$Co$_2$O$_7$, and hybridized bonding and anti-bonding molecular Wannier orbitals of (b) $d_{z^2}-p_z-d_{z^2}$ $\sigma$ bonds, (c) $d_{xz}-p_x-d_{xz}$ $\pi$ bonds and (d) $d_{yz}-p_y-d_{yz}$ $\pi$ bonds, imposed by interlayer couplings. The molecular orbitals, starting from bilayer Co$^{4+}$ ($3d^5, t_{2g}^3e_g^2$) and Co$^{4+}$ ($3d^7\underline L^2$) states, are firstly formed by head-to-head hopping via non-zero overlap of $d_{z^2}$ and $p_z$ orbitals, or by shoulder-by-shoulder hopping via non-zero overlap of $d_{xz}$ and $p_x$ ($d_{yz}$ and $p_y$) orbitals, and further develop antiferromagnetic tendency by invoking the on-site Hubbard interaction (in the strong coupling limit) during the hopping process, leading to half-filling of $d_{z^2}$, $d_{xz}$ and $d_{yz}$ orbitals. Notice that for bonding orbitals the energy splitting due to the $p-d$ hybridization is opposite to the regular octahedral crystal field, where $e_g$ levels lie below $t_{2g}$ multiplets. The remaining $d_{x^2-y^2}$ and $d_{xy}$ orbitals are lifted by octahedral crystal field. The charge disproportionation between $d_{x^2-y^2}$ and $d_{xy}$ orbitals are represented by the generation of a hole (in red circle) within the $d_{xy}$ orbitals and an electron with spin $\uparrow$ within the $d_{x^2-y^2}$ orbitals through possible charge transfer via $p$ orbitals.}
\label{energy_diagram}
\end{figure}

Now, the electronic structure illustrated in Fig.\ref{band_structure}(b-d) can be well interpreted by formally considering molecular orbitals that originate from Co$^{4+}$ ($3d^5$) and Co$^{4+}$ ($3d^7\underline L^2$) states within the bilayer structure, taking into account the strong interlayer orbital hybridization and charge disproportionation. The interlayer molecular orbital energy diagram is shown in Fig.\ref{energy_diagram}(a). Firstly, strong interlayer coupling leads to significant energy splitting of the $d_{z^2}$ and $d_{xz}/d_{yz}$ orbitals in Sr$_3$Co$_2$O$_7$. The $d_{z^2}$ orbitals exhibit strong hybridization with the inter-layer apical O $p_z$ orbitals, which form $\sigma$-bonding and antibonding molecular orbitals, while the $d_{xz}/d_{yz}$ orbitals hybridize with the inner-layer apical O $p_x$/$p_y$ orbitals, forming $\pi$-bonding and antibonding molecular orbitals. The molecular Wannier functions in Fig.\ref{energy_diagram}(b) clearly show the $\sigma$-bonding character via head-to-head overlaps of Co $d_{z^2}$ and O $p_z$ orbitals, while molecular Wannier functions in Fig.\ref{energy_diagram}(c-d) clearly show the $\pi$-bonding character via shoulder-by-shoulder overlaps of Co $d_{xy}$ and O $p_x$ orbitals as well as Co $d_{yz}$ and O $p_y$ orbitals (see construction of molecular Wannier functions in section B in Supplementary Materials). Secondly, the energy levels of remaining $d_{x^2-y^2}$ and $d_{xy}$ orbitals are lifted by point-charge octahedral crystal field. They further develop charge disproportionation with delocalized behavior. The charge fluctuation between $d_{x^2-y^2}$ and $d_{xy}$ orbitals are represented in Fig.\ref{energy_diagram}(a) by the generation of a hole (in red circle) within the $d_{xy}$ orbitals and an electron with spin $\uparrow$ within the $d_{x^2-y^2}$ orbitals through possible charge transfer via $p$ orbitals.

The A-type antiferromagnetic polar metallicity in Sr$_3$Co$_2$O$_7$ could thus be well understood within the molecular orbital with charge disproportionation framework. Electrons in $d_{z^2}$ and $d_{xz}/d_{yz}$ orbitals become fully localized favoring strong antiferromagnetic super-exchange interactions according to the Goodenough-Kanamori rule, and antiferromagnetic tendency is strongest in the $d_{z^2}$ orbitals due to the nature of $\sigma$ bond. And the delocalized $d_{x^2-y^2}$ and $d_{xy}$ electrons gives rise to the polar metallicity from their charge disproportionation. When these molecular orbitals develop into bands, cobaltate Sr$_3$Co$_2$O$_7$ can be viewed as molecular orbital crystal formed by two-dimensional periodic arrangement of these Co-O-Co units. Also, the polar antiferromagnetic metallicity driven by the charge disproportionation in negative charge transfer gap regime, is an analog of joint $\mathcal{PT}$ symmetry breaking in topological insulators with negative mass term. Our formulation of the molecular orbitals consistently unifies the polarity, metallicity, and the antiferromagnetism in Sr$_3$Co$_2$O$_7$ through the coexistentence of localized and delocalized Co-$d$ electrons. 

\textit{Effective model Hamiltonian}--
The charge disproportionation scenario permits consequent conducting carriers to couple with interlayer local spins via Hund's coupling, therefore our proposed theory can be casted into an effective model Hamiltonian which describes all features of the A-type antiferromagnetic polar metallic molecular orbital crystal. We start with Co$^{4+}$ 3$d^6 \underline L$ on the cobalt bilayer structure with interlayer AF superexchange interactions,
\begin{equation}
\mathcal{H}_{\mathrm{AF}}=J_{\mathrm{AF}} \sum_{i} \hat{S}_{1,i} \hat{S}_{2,i}
\end{equation}
Here $\hat{S}_{l,i}$ ($l=1,2$ for layer index) denotes localized spins from $d_{z^2}$, $d_{xz}$ and $d_{yz}$ orbitals at the same atomic position $i$ within each layer. These localized spins carry maximum $S=\frac{3}{2}$ quantum number according to the Hund's rule. Adopting the molecular crystal picture allows us to omit the layer index and focus solely on the site index $i$. Due to the charge disproportionation, there are conducting $d_{xy}$ holons and $d_{x^2-y^2}$ electrons relative to the Co$^{4+}$ 3$d^6 \underline L$ within the Co-Co dimers. This charge disproportionation can be phenomenonally described by the chemical potential difference $\Delta \mu=\epsilon_{d_{xy}}-\epsilon_{d_{x^2-y^2}} $ between $d_{xy}$ and $d_{x^2-y^2}$ orbitals, which is microscopically governed by Hund's rule as shown above. We further consider the Hund's coupling between localized spins and these conducting $d_{xy}$ and $d_{x^2-y^2}$ carriers. This leads to the following double-exchange Hamiltonian:
\begin{eqnarray}
\mathcal{H}_{\mathrm{DE}}&=&-t \sum_{\langle i j\rangle,\alpha, \sigma} c_{i \sigma \alpha}^{\dagger} c_{j \sigma \alpha}-t' \sum_{\langle i j\rangle,\alpha, \alpha' \sigma} c_{i \sigma \alpha}^{\dagger} c_{j \sigma \alpha'}\notag \\
&&-J_{\mathrm{H}} \sum_{i,\alpha, \sigma} \hat{S}_{i} \cdot c_{i \sigma \alpha}^{\dagger} \hat{\boldsymbol{\sigma}} c_{i \sigma \alpha}-\sum_{i, \alpha \sigma} \epsilon_\alpha \cdot n_{i \sigma \alpha}
\end{eqnarray}
where $c_{i \sigma \alpha}^{\dagger}$ and $c_{i\sigma \alpha}$ represent the creation and annihilation operators for conducting electrons in $\alpha$ orbitals ($\alpha$=$d_{xy}$ or $d_{x^2-y^2}$) and $\sigma$ spin species, with in-plane inter-dimer hopping parameters $t$ and $t'$ for hopping within the same and different orbitals, respectively.
For brevity, we take electron hopping integral for $d_{xy}$ and $d_{x^2-y^2}$ orbitals as equal. Actually, the band width for $d_{xy}$ orbital is much smaller than that of $d_{x^2-y^2}$ orbital near the fermi level, which indicates hopping integral of $d_{xy}$ orbitals is strongly renormalized. $J_H$ term is the Hund's rule exchange between localized spin $\hat{S}_i$ with conducting electrons carrying spin $\hat{{\sigma}}$.
It favors the in-plane ferromagentic correlations by gaining the kinetic and Hund's energy from electrons with equal spins.

\begin{figure}[t]
\includegraphics[width=0.42\textwidth]{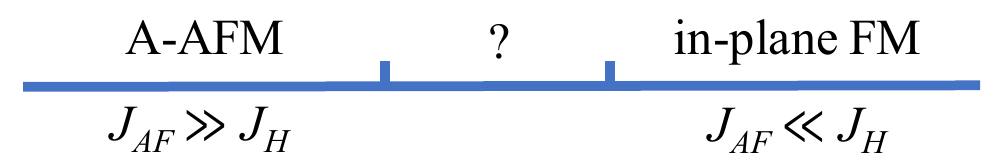}
\caption{Schematic phase diagram of the effective model Hamiltonian on cobalt bilayer structure in the $J_{AF}$ versus $J_H$ axis ($\Delta \mu \neq$ 0). }
\label{phase_diagram}
\end{figure}

All together, the total model Hamiltonian for cobaltate Sr$_3$Co$_2$O$_7$
consists of two terms
\begin{equation}
\mathcal{H}=\mathcal{H}_{\mathrm{AF}}+\mathcal{H}_{\mathrm{DE}}
\end{equation}
There are several key energy scales: $t(t')$, $\Delta \mu$, $J_{AF}$ and $J_H$. The Heisenberg superexchange $J_{AF}$ accounts for the interlayer antiferromagnetism from $d_{z^2}$ and $d_{xz}/d_{yz}$ orbitals bridging by apical oxygen $p$ orbitals. Another energy scale Hund's coupling $J_H$, which couples interlayer local spins and conducting carriers, facilitates the in-plane ferromagnetic order. The chemical potential $\Delta \mu$ determines the population number of $d_{xy}$ and $d_{x^2-y^2}$ orbital, mimicking the charge disproportionation and resultant polarity. The combination of charge disproportionation and inter-dimer hopping of delocalized $d_{xy}$ and $d_{x^2-y^2}$ orbitals is the source of metallity. Without solving the ground state of the model Hamiltonian, we can discuss some limits in the parameter space. When the Heisenberg superexchange $J_{AF}$ dominates (i.e., $J_{AF} \gg J_H$), facilitating antiparallel alignment of neighboring spins between layers, the system falls into the A-type antiferromagnetism (A-AFM) phase. When the Hund's coupling $J_H$ dominates (i.e., $J_H \gg J_{AF}$), the strong Hund’s coupling aligns the itinerant and local spins and promotes parallel alignment within the plane through the double-exchange mechanism, ultimately the system is a in-plane ferromagnetic metal (in-plane FM, see a mathematical proof in section F in Supplementary Materials). And the polar antiferromagnetic metallic state just lies in the crossover regime between the interlayer AFM and in-plane FM orders in the phase diagram. The schematic phase diagram is summarized in the Fig.\ref{phase_diagram}.

\textit{Discussion and conclusion}-- It is interesting to compare the origins of polarity of Sr$_{3}$Co$_{2}$O$_{7}$ with that of Ca$_{3}$Ru$_{2}$O$_{7}$. Different from the charge disproportion mechanism in Sr$_{3}$Co$_{2}$O$_{7}$, Ca$_{3}$Ru$_{2}$O$_{7}$ exhibits polarity through a structurally driven octahedral rotation and tilt mechanism, where the rotation ($X_2^+$ mode) and tilt ($X_3^-$ mode) of adjacent oxygen octahedra in the bilayer structrue triggers polar phonon instability ($\Gamma_5^-$ mode)\cite{sokolov2019metamagnetic}. The stability of the polarity comes from a trilinear anharmonic coupling of the form $\sim Q_{\Gamma_5^-} Q_{X_2^+} Q_{X_3^-}$, whereby the energy gain from their mutual coexistence surpasses that obtained from any individual mode\cite{PhysRevLett.106.107204}. Given that Ru$^{4+}$ has the same nominal valence as Co$^{4+}$, the contrasting polarity origins might be related to the different orbital occupations of Ru$^{4+}$ and radii of the A-site cations. Ru$^{4+}$ is in the 3$d^4$ valence configuration which suggests a different spin state, and the cation size mismatch helps stabilize octahedral rotations and tilts to lower the electrostatic energy, thus triggering the polarity inside the plane in Ca$_{3}$Ru$_{2}$O$_{7}$.

In conclusion, we propose a microscopic charge disproportionation mechanism underlying the synergistic emergence of metallicity, polarity, and antiferromagnetism in the bilayer Ruddlesden-Popper perovskite Sr$_3$Co$_2$O$_7$ through first-principles DFT calculations and molecular orbital analyses. It essentially establishes an orbital-selective electronic behavior of coexisting localized and delocalized states in multi-orbital quantum-confined system as a novel magnetoelectric coupling scheme, where polarized conducting carriers interact with localized spins in molecular spin dimers via Hund’s coupling, enabling A-type AFM ordering with in-plane double-exchange ferromagnetic components. Our molecular orbital formulation not only reconciles the incompatibility of metallicity, polarity, and antiferromagnetism in Sr$_3$Co$_2$O$_7$, enabling the construction of effective model Hamitonian, but also establishes a universal paradigm of orbital selective physics related to double-layered RP perovskite oxides.

\textit{Acknowledgments}-- This work  was supported by the National Key Research and Development Program of MOST of China (Grant No. 2023YFA1406400 and No. 2022YFA1402703) and the NSF of China (Grants No. 52025024, No. 52388201, No. 12574271 and 92365204), Sino-German Mobility program (No. M-0006), and Shanghai 2021- Fundamental Research Area (No. 21JC1404700). Part of the calculations was performed at the HPC Platform of ShanghaiTech University Library and Information Services, and the School of Physical Science and Technology.

\bibliography{ref}

\begin{thebibliography}{31}%
\makeatletter
\providecommand \@ifxundefined [1]{%
 \@ifx{#1\undefined}
}%
\providecommand \@ifnum [1]{%
 \ifnum #1\expandafter \@firstoftwo
 \else \expandafter \@secondoftwo
 \fi
}%
\providecommand \@ifx [1]{%
 \ifx #1\expandafter \@firstoftwo
 \else \expandafter \@secondoftwo
 \fi
}%
\providecommand \natexlab [1]{#1}%
\providecommand \enquote  [1]{``#1''}%
\providecommand \bibnamefont  [1]{#1}%
\providecommand \bibfnamefont [1]{#1}%
\providecommand \citenamefont [1]{#1}%
\providecommand \href@noop [0]{\@secondoftwo}%
\providecommand \href [0]{\begingroup \@sanitize@url \@href}%
\providecommand \@href[1]{\@@startlink{#1}\@@href}%
\providecommand \@@href[1]{\endgroup#1\@@endlink}%
\providecommand \@sanitize@url [0]{\catcode `\\12\catcode `\$12\catcode
  `\&12\catcode `\#12\catcode `\^12\catcode `\_12\catcode `\%12\relax}%
\providecommand \@@startlink[1]{}%
\providecommand \@@endlink[0]{}%
\providecommand \url  [0]{\begingroup\@sanitize@url \@url }%
\providecommand \@url [1]{\endgroup\@href {#1}{\urlprefix }}%
\providecommand \urlprefix  [0]{URL }%
\providecommand \Eprint [0]{\href }%
\providecommand \doibase [0]{https://doi.org/}%
\providecommand \selectlanguage [0]{\@gobble}%
\providecommand \bibinfo  [0]{\@secondoftwo}%
\providecommand \bibfield  [0]{\@secondoftwo}%
\providecommand \translation [1]{[#1]}%
\providecommand \BibitemOpen [0]{}%
\providecommand \bibitemStop [0]{}%
\providecommand \bibitemNoStop [0]{.\EOS\space}%
\providecommand \EOS [0]{\spacefactor3000\relax}%
\providecommand \BibitemShut  [1]{\csname bibitem#1\endcsname}%
\let\auto@bib@innerbib\@empty
\bibitem [{\citenamefont {Princep}\ \emph {et~al.}(2020)\citenamefont
  {Princep}, \citenamefont {Feng}, \citenamefont {Guo}, \citenamefont {Lang},
  \citenamefont {Weng}, \citenamefont {Manuel}, \citenamefont {Khalyavin},
  \citenamefont {Senyshyn}, \citenamefont {Rahn}, \citenamefont {Yuan},
  \citenamefont {Matsushita}, \citenamefont {Blundell}, \citenamefont
  {Yamaura},\ and\ \citenamefont {Boothroyd}}]{PhysRevB.102.104410}%
  \BibitemOpen
  \bibfield  {author} {\bibinfo {author} {\bibfnamefont {A.~J.}\ \bibnamefont
  {Princep}}, \bibinfo {author} {\bibfnamefont {H.~L.}\ \bibnamefont {Feng}},
  \bibinfo {author} {\bibfnamefont {Y.~F.}\ \bibnamefont {Guo}}, \bibinfo
  {author} {\bibfnamefont {F.}~\bibnamefont {Lang}}, \bibinfo {author}
  {\bibfnamefont {H.~M.}\ \bibnamefont {Weng}}, \bibinfo {author}
  {\bibfnamefont {P.}~\bibnamefont {Manuel}}, \bibinfo {author} {\bibfnamefont
  {D.}~\bibnamefont {Khalyavin}}, \bibinfo {author} {\bibfnamefont
  {A.}~\bibnamefont {Senyshyn}}, \bibinfo {author} {\bibfnamefont {M.~C.}\
  \bibnamefont {Rahn}}, \bibinfo {author} {\bibfnamefont {Y.~H.}\ \bibnamefont
  {Yuan}}, \bibinfo {author} {\bibfnamefont {Y.}~\bibnamefont {Matsushita}},
  \bibinfo {author} {\bibfnamefont {S.~J.}\ \bibnamefont {Blundell}}, \bibinfo
  {author} {\bibfnamefont {K.}~\bibnamefont {Yamaura}},\ and\ \bibinfo {author}
  {\bibfnamefont {A.~T.}\ \bibnamefont {Boothroyd}},\ }\bibfield  {title}
  {\bibinfo {title} {Magnetically driven loss of centrosymmetry in metallic
  {${\mathrm{Pb}}_{2}{\mathrm{CoOsO}}_{6}$}},\ }\href
  {https://doi.org/10.1103/PhysRevB.102.104410} {\bibfield  {journal} {\bibinfo
   {journal} {Phys. Rev. B}\ }\textbf {\bibinfo {volume} {102}},\ \bibinfo
  {pages} {104410} (\bibinfo {year} {2020})}\BibitemShut {NoStop}%
\bibitem [{\citenamefont {Lei}\ \emph {et~al.}(2018)\citenamefont {Lei},
  \citenamefont {Gu}, \citenamefont {Puggioni}, \citenamefont {Stone},
  \citenamefont {Peng}, \citenamefont {Ge}, \citenamefont {Wang}, \citenamefont
  {Wang}, \citenamefont {Yuan}, \citenamefont {Wang}, \citenamefont {Mao},
  \citenamefont {Rondinelli},\ and\ \citenamefont
  {Gopalan}}]{lei2018observation}%
  \BibitemOpen
  \bibfield  {author} {\bibinfo {author} {\bibfnamefont {S.}~\bibnamefont
  {Lei}}, \bibinfo {author} {\bibfnamefont {M.}~\bibnamefont {Gu}}, \bibinfo
  {author} {\bibfnamefont {D.}~\bibnamefont {Puggioni}}, \bibinfo {author}
  {\bibfnamefont {G.}~\bibnamefont {Stone}}, \bibinfo {author} {\bibfnamefont
  {J.}~\bibnamefont {Peng}}, \bibinfo {author} {\bibfnamefont {J.}~\bibnamefont
  {Ge}}, \bibinfo {author} {\bibfnamefont {Y.}~\bibnamefont {Wang}}, \bibinfo
  {author} {\bibfnamefont {B.}~\bibnamefont {Wang}}, \bibinfo {author}
  {\bibfnamefont {Y.}~\bibnamefont {Yuan}}, \bibinfo {author} {\bibfnamefont
  {K.}~\bibnamefont {Wang}}, \bibinfo {author} {\bibfnamefont {Z.}~\bibnamefont
  {Mao}}, \bibinfo {author} {\bibfnamefont {J.~M.}\ \bibnamefont
  {Rondinelli}},\ and\ \bibinfo {author} {\bibfnamefont {V.}~\bibnamefont
  {Gopalan}},\ }\bibfield  {title} {\bibinfo {title} {Observation of
  quasi-two-dimensional polar domains and ferroelastic switching in a metal,
  {$\mathrm{Ca}_3\mathrm{Ru}_2\mathrm{O}_7$}},\ }\href
  {https://doi.org/10.1021/acs.nanolett.8b00633} {\bibfield  {journal}
  {\bibinfo  {journal} {Nano Letters}\ }\textbf {\bibinfo {volume} {18}},\
  \bibinfo {pages} {3088–3095} (\bibinfo {year} {2018})}\BibitemShut
  {NoStop}%
\bibitem [{\citenamefont {\ifmmode~\check{S}\else \v{S}\fi{}mejkal}\ \emph
  {et~al.}(2022)\citenamefont {\ifmmode~\check{S}\else \v{S}\fi{}mejkal},
  \citenamefont {Sinova},\ and\ \citenamefont
  {Jungwirth}}]{PhysRevX.12.031042}%
  \BibitemOpen
  \bibfield  {author} {\bibinfo {author} {\bibfnamefont {L.}~\bibnamefont
  {\ifmmode~\check{S}\else \v{S}\fi{}mejkal}}, \bibinfo {author} {\bibfnamefont
  {J.}~\bibnamefont {Sinova}},\ and\ \bibinfo {author} {\bibfnamefont
  {T.}~\bibnamefont {Jungwirth}},\ }\bibfield  {title} {\bibinfo {title}
  {Beyond conventional ferromagnetism and antiferromagnetism: A phase with
  nonrelativistic spin and crystal rotation symmetry},\ }\href
  {https://doi.org/10.1103/PhysRevX.12.031042} {\bibfield  {journal} {\bibinfo
  {journal} {Phys. Rev. X}\ }\textbf {\bibinfo {volume} {12}},\ \bibinfo
  {pages} {031042} (\bibinfo {year} {2022})}\BibitemShut {NoStop}%
\bibitem [{\citenamefont {Song}\ \emph {et~al.}(2023)\citenamefont {Song},
  \citenamefont {Doyle}, \citenamefont {Pan}, \citenamefont {El~Baggari},
  \citenamefont {Ferenc~Segedin}, \citenamefont {Córdova~Carrizales},
  \citenamefont {Nordlander}, \citenamefont {Tzschaschel}, \citenamefont
  {Ehrets}, \citenamefont {Hasan}, \citenamefont {El-Sherif}, \citenamefont
  {Krishna}, \citenamefont {Hanson}, \citenamefont {LaBollita}, \citenamefont
  {Bostwick}, \citenamefont {Jozwiak}, \citenamefont {Rotenberg}, \citenamefont
  {Xu}, \citenamefont {Lanzara}, \citenamefont {N’Diaye}, \citenamefont
  {Heikes}, \citenamefont {Liu}, \citenamefont {Paik}, \citenamefont {Brooks},
  \citenamefont {Pamuk}, \citenamefont {Heron}, \citenamefont {Shafer},
  \citenamefont {Ratcliff}, \citenamefont {Botana}, \citenamefont
  {Moreschini},\ and\ \citenamefont {Mundy}}]{Song2023}%
  \BibitemOpen
  \bibfield  {author} {\bibinfo {author} {\bibfnamefont {Q.}~\bibnamefont
  {Song}}, \bibinfo {author} {\bibfnamefont {S.}~\bibnamefont {Doyle}},
  \bibinfo {author} {\bibfnamefont {G.~A.}\ \bibnamefont {Pan}}, \bibinfo
  {author} {\bibfnamefont {I.}~\bibnamefont {El~Baggari}}, \bibinfo {author}
  {\bibfnamefont {D.}~\bibnamefont {Ferenc~Segedin}}, \bibinfo {author}
  {\bibfnamefont {D.}~\bibnamefont {Córdova~Carrizales}}, \bibinfo {author}
  {\bibfnamefont {J.}~\bibnamefont {Nordlander}}, \bibinfo {author}
  {\bibfnamefont {C.}~\bibnamefont {Tzschaschel}}, \bibinfo {author}
  {\bibfnamefont {J.~R.}\ \bibnamefont {Ehrets}}, \bibinfo {author}
  {\bibfnamefont {Z.}~\bibnamefont {Hasan}}, \bibinfo {author} {\bibfnamefont
  {H.}~\bibnamefont {El-Sherif}}, \bibinfo {author} {\bibfnamefont
  {J.}~\bibnamefont {Krishna}}, \bibinfo {author} {\bibfnamefont
  {C.}~\bibnamefont {Hanson}}, \bibinfo {author} {\bibfnamefont
  {H.}~\bibnamefont {LaBollita}}, \bibinfo {author} {\bibfnamefont
  {A.}~\bibnamefont {Bostwick}}, \bibinfo {author} {\bibfnamefont
  {C.}~\bibnamefont {Jozwiak}}, \bibinfo {author} {\bibfnamefont
  {E.}~\bibnamefont {Rotenberg}}, \bibinfo {author} {\bibfnamefont {S.-Y.}\
  \bibnamefont {Xu}}, \bibinfo {author} {\bibfnamefont {A.}~\bibnamefont
  {Lanzara}}, \bibinfo {author} {\bibfnamefont {A.~T.}\ \bibnamefont
  {N’Diaye}}, \bibinfo {author} {\bibfnamefont {C.~A.}\ \bibnamefont
  {Heikes}}, \bibinfo {author} {\bibfnamefont {Y.}~\bibnamefont {Liu}},
  \bibinfo {author} {\bibfnamefont {H.}~\bibnamefont {Paik}}, \bibinfo {author}
  {\bibfnamefont {C.~M.}\ \bibnamefont {Brooks}}, \bibinfo {author}
  {\bibfnamefont {B.}~\bibnamefont {Pamuk}}, \bibinfo {author} {\bibfnamefont
  {J.~T.}\ \bibnamefont {Heron}}, \bibinfo {author} {\bibfnamefont
  {P.}~\bibnamefont {Shafer}}, \bibinfo {author} {\bibfnamefont {W.~D.}\
  \bibnamefont {Ratcliff}}, \bibinfo {author} {\bibfnamefont {A.~S.}\
  \bibnamefont {Botana}}, \bibinfo {author} {\bibfnamefont {L.}~\bibnamefont
  {Moreschini}},\ and\ \bibinfo {author} {\bibfnamefont {J.~A.}\ \bibnamefont
  {Mundy}},\ }\bibfield  {title} {\bibinfo {title} {Antiferromagnetic metal
  phase in an electron-doped rare-earth nickelate},\ }\href
  {https://doi.org/10.1038/s41567-022-01907-2} {\bibfield  {journal} {\bibinfo
  {journal} {Nature Physics}\ }\textbf {\bibinfo {volume} {19}},\ \bibinfo
  {pages} {522–528} (\bibinfo {year} {2023})}\BibitemShut {NoStop}%
\bibitem [{\citenamefont {Cao}\ \emph {et~al.}(1997)\citenamefont {Cao},
  \citenamefont {McCall}, \citenamefont {Crow},\ and\ \citenamefont
  {Guertin}}]{PhysRevLett.78.1751}%
  \BibitemOpen
  \bibfield  {author} {\bibinfo {author} {\bibfnamefont {G.}~\bibnamefont
  {Cao}}, \bibinfo {author} {\bibfnamefont {S.}~\bibnamefont {McCall}},
  \bibinfo {author} {\bibfnamefont {J.~E.}\ \bibnamefont {Crow}},\ and\
  \bibinfo {author} {\bibfnamefont {R.~P.}\ \bibnamefont {Guertin}},\
  }\bibfield  {title} {\bibinfo {title} {Observation of a metallic
  antiferromagnetic phase and metal to nonmetal transition in
  {${\mathrm{Ca}}_{3}{\mathrm{Ru}}_{2}\mathrm{O}_{7}$}},\ }\href
  {https://doi.org/10.1103/PhysRevLett.78.1751} {\bibfield  {journal} {\bibinfo
   {journal} {Phys. Rev. Lett.}\ }\textbf {\bibinfo {volume} {78}},\ \bibinfo
  {pages} {1751} (\bibinfo {year} {1997})}\BibitemShut {NoStop}%
\bibitem [{\citenamefont {Anderson}\ and\ \citenamefont
  {Blount}(1965)}]{anderson1965symmetry}%
  \BibitemOpen
  \bibfield  {author} {\bibinfo {author} {\bibfnamefont {P.~W.}\ \bibnamefont
  {Anderson}}\ and\ \bibinfo {author} {\bibfnamefont {E.}~\bibnamefont
  {Blount}},\ }\bibfield  {title} {\bibinfo {title} {Symmetry considerations on
  martensitic transformations:" ferroelectric" metals?},\ }\href@noop {}
  {\bibfield  {journal} {\bibinfo  {journal} {Physical Review Letters}\
  }\textbf {\bibinfo {volume} {14}},\ \bibinfo {pages} {217} (\bibinfo {year}
  {1965})}\BibitemShut {NoStop}%
\bibitem [{\citenamefont {Kim}\ \emph {et~al.}(2016)\citenamefont {Kim},
  \citenamefont {Puggioni}, \citenamefont {Yuan}, \citenamefont {Xie},
  \citenamefont {Zhou}, \citenamefont {Campbell}, \citenamefont {Ryan},
  \citenamefont {Choi}, \citenamefont {Kim}, \citenamefont {Patzner},
  \citenamefont {Ryu}, \citenamefont {Podkaminer}, \citenamefont {Irwin},
  \citenamefont {Ma}, \citenamefont {Fennie}, \citenamefont {Rzchowski},
  \citenamefont {Pan}, \citenamefont {Gopalan}, \citenamefont {Rondinelli},\
  and\ \citenamefont {Eom}}]{kim2016polar}%
  \BibitemOpen
  \bibfield  {author} {\bibinfo {author} {\bibfnamefont {T.}~\bibnamefont
  {Kim}}, \bibinfo {author} {\bibfnamefont {D.}~\bibnamefont {Puggioni}},
  \bibinfo {author} {\bibfnamefont {Y.}~\bibnamefont {Yuan}}, \bibinfo {author}
  {\bibfnamefont {L.}~\bibnamefont {Xie}}, \bibinfo {author} {\bibfnamefont
  {H.}~\bibnamefont {Zhou}}, \bibinfo {author} {\bibfnamefont {N.}~\bibnamefont
  {Campbell}}, \bibinfo {author} {\bibfnamefont {P.}~\bibnamefont {Ryan}},
  \bibinfo {author} {\bibfnamefont {Y.}~\bibnamefont {Choi}}, \bibinfo {author}
  {\bibfnamefont {J.-W.}\ \bibnamefont {Kim}}, \bibinfo {author} {\bibfnamefont
  {J.}~\bibnamefont {Patzner}}, \bibinfo {author} {\bibfnamefont
  {S.}~\bibnamefont {Ryu}}, \bibinfo {author} {\bibfnamefont {J.}~\bibnamefont
  {Podkaminer}}, \bibinfo {author} {\bibfnamefont {J.}~\bibnamefont {Irwin}},
  \bibinfo {author} {\bibfnamefont {Y.}~\bibnamefont {Ma}}, \bibinfo {author}
  {\bibfnamefont {C.}~\bibnamefont {Fennie}}, \bibinfo {author} {\bibfnamefont
  {M.}~\bibnamefont {Rzchowski}}, \bibinfo {author} {\bibfnamefont
  {X.}~\bibnamefont {Pan}}, \bibinfo {author} {\bibfnamefont {V.}~\bibnamefont
  {Gopalan}}, \bibinfo {author} {\bibfnamefont {J.~M.}\ \bibnamefont
  {Rondinelli}},\ and\ \bibinfo {author} {\bibfnamefont {C.}~\bibnamefont
  {Eom}},\ }\bibfield  {title} {\bibinfo {title} {Polar metals by geometric
  design},\ }\href@noop {} {\bibfield  {journal} {\bibinfo  {journal} {Nature}\
  }\textbf {\bibinfo {volume} {533}},\ \bibinfo {pages} {68} (\bibinfo {year}
  {2016})}\BibitemShut {NoStop}%
\bibitem [{\citenamefont {Bhowal}\ and\ \citenamefont
  {Spaldin}(2023)}]{bhowal2023polar}%
  \BibitemOpen
  \bibfield  {author} {\bibinfo {author} {\bibfnamefont {S.}~\bibnamefont
  {Bhowal}}\ and\ \bibinfo {author} {\bibfnamefont {N.~A.}\ \bibnamefont
  {Spaldin}},\ }\bibfield  {title} {\bibinfo {title} {Polar metals: principles
  and prospects},\ }\href@noop {} {\bibfield  {journal} {\bibinfo  {journal}
  {Annual Review of Materials Research}\ }\textbf {\bibinfo {volume} {53}},\
  \bibinfo {pages} {53} (\bibinfo {year} {2023})}\BibitemShut {NoStop}%
\bibitem [{\citenamefont {Salamon}\ and\ \citenamefont
  {Jaime}(2001)}]{RevModPhys.73.583}%
  \BibitemOpen
  \bibfield  {author} {\bibinfo {author} {\bibfnamefont {M.~B.}\ \bibnamefont
  {Salamon}}\ and\ \bibinfo {author} {\bibfnamefont {M.}~\bibnamefont
  {Jaime}},\ }\bibfield  {title} {\bibinfo {title} {The physics of manganites:
  Structure and transport},\ }\href {https://doi.org/10.1103/RevModPhys.73.583}
  {\bibfield  {journal} {\bibinfo  {journal} {Rev. Mod. Phys.}\ }\textbf
  {\bibinfo {volume} {73}},\ \bibinfo {pages} {583} (\bibinfo {year}
  {2001})}\BibitemShut {NoStop}%
\bibitem [{\citenamefont {Benedek}\ and\ \citenamefont
  {Fennie}(2011)}]{PhysRevLett.106.107204}%
  \BibitemOpen
  \bibfield  {author} {\bibinfo {author} {\bibfnamefont {N.~A.}\ \bibnamefont
  {Benedek}}\ and\ \bibinfo {author} {\bibfnamefont {C.~J.}\ \bibnamefont
  {Fennie}},\ }\bibfield  {title} {\bibinfo {title} {Hybrid improper
  ferroelectricity: A mechanism for controllable polarization-magnetization
  coupling},\ }\href {https://doi.org/10.1103/PhysRevLett.106.107204}
  {\bibfield  {journal} {\bibinfo  {journal} {Phys. Rev. Lett.}\ }\textbf
  {\bibinfo {volume} {106}},\ \bibinfo {pages} {107204} (\bibinfo {year}
  {2011})}\BibitemShut {NoStop}%
\bibitem [{\citenamefont {Pitcher}\ \emph {et~al.}(2015)\citenamefont
  {Pitcher}, \citenamefont {Mandal}, \citenamefont {Dyer}, \citenamefont
  {Alaria}, \citenamefont {Borisov}, \citenamefont {Niu}, \citenamefont
  {Claridge},\ and\ \citenamefont {Rosseinsky}}]{doi:10.1126/science.1262118}%
  \BibitemOpen
  \bibfield  {author} {\bibinfo {author} {\bibfnamefont {M.~J.}\ \bibnamefont
  {Pitcher}}, \bibinfo {author} {\bibfnamefont {P.}~\bibnamefont {Mandal}},
  \bibinfo {author} {\bibfnamefont {M.~S.}\ \bibnamefont {Dyer}}, \bibinfo
  {author} {\bibfnamefont {J.}~\bibnamefont {Alaria}}, \bibinfo {author}
  {\bibfnamefont {P.}~\bibnamefont {Borisov}}, \bibinfo {author} {\bibfnamefont
  {H.}~\bibnamefont {Niu}}, \bibinfo {author} {\bibfnamefont {J.~B.}\
  \bibnamefont {Claridge}},\ and\ \bibinfo {author} {\bibfnamefont {M.~J.}\
  \bibnamefont {Rosseinsky}},\ }\bibfield  {title} {\bibinfo {title} {Tilt
  engineering of spontaneous polarization and magnetization above 300 k in a
  bulk layered perovskite},\ }\href {https://doi.org/10.1126/science.1262118}
  {\bibfield  {journal} {\bibinfo  {journal} {Science}\ }\textbf {\bibinfo
  {volume} {347}},\ \bibinfo {pages} {420} (\bibinfo {year}
  {2015})}\BibitemShut {NoStop}%
\bibitem [{\citenamefont {Smith}\ \emph {et~al.}(2019)\citenamefont {Smith},
  \citenamefont {Nowadnick}, \citenamefont {Fan}, \citenamefont {Khatib},
  \citenamefont {Lim}, \citenamefont {Gao}, \citenamefont {Harms},
  \citenamefont {Neal}, \citenamefont {Kirkland}, \citenamefont {Martin},
  \citenamefont {Won}, \citenamefont {Raschke}, \citenamefont {Cheong},
  \citenamefont {Fennie}, \citenamefont {Carr}, \citenamefont {Bechtel},\ and\
  \citenamefont {Musfeldt}}]{smith2019infrared}%
  \BibitemOpen
  \bibfield  {author} {\bibinfo {author} {\bibfnamefont {K.~A.}\ \bibnamefont
  {Smith}}, \bibinfo {author} {\bibfnamefont {E.~A.}\ \bibnamefont
  {Nowadnick}}, \bibinfo {author} {\bibfnamefont {S.}~\bibnamefont {Fan}},
  \bibinfo {author} {\bibfnamefont {O.}~\bibnamefont {Khatib}}, \bibinfo
  {author} {\bibfnamefont {S.~J.}\ \bibnamefont {Lim}}, \bibinfo {author}
  {\bibfnamefont {B.}~\bibnamefont {Gao}}, \bibinfo {author} {\bibfnamefont
  {N.~C.}\ \bibnamefont {Harms}}, \bibinfo {author} {\bibfnamefont {S.~N.}\
  \bibnamefont {Neal}}, \bibinfo {author} {\bibfnamefont {J.~K.}\ \bibnamefont
  {Kirkland}}, \bibinfo {author} {\bibfnamefont {M.~C.}\ \bibnamefont
  {Martin}}, \bibinfo {author} {\bibfnamefont {C.~J.}\ \bibnamefont {Won}},
  \bibinfo {author} {\bibfnamefont {M.~B.}\ \bibnamefont {Raschke}}, \bibinfo
  {author} {\bibfnamefont {S.-W.}\ \bibnamefont {Cheong}}, \bibinfo {author}
  {\bibfnamefont {C.~J.}\ \bibnamefont {Fennie}}, \bibinfo {author}
  {\bibfnamefont {G.~L.}\ \bibnamefont {Carr}}, \bibinfo {author}
  {\bibfnamefont {H.~A.}\ \bibnamefont {Bechtel}},\ and\ \bibinfo {author}
  {\bibfnamefont {J.~L.}\ \bibnamefont {Musfeldt}},\ }\bibfield  {title}
  {\bibinfo {title} {Infrared nano-spectroscopy of ferroelastic domain walls in
  hybrid improper ferroelectric {$\mathrm{Ca}_3\mathrm{Ti}_2\mathrm{O}_7$}},\
  }\bibfield  {journal} {\bibinfo  {journal} {Nature Communications}\ }\textbf
  {\bibinfo {volume} {10}},\ \href {https://doi.org/10.1038/s41467-019-13066-9}
  {10.1038/s41467-019-13066-9} (\bibinfo {year} {2019})\BibitemShut {NoStop}%
\bibitem [{\citenamefont {Lee}\ \emph {et~al.}(2013)\citenamefont {Lee},
  \citenamefont {Orloff}, \citenamefont {Birol}, \citenamefont {Zhu},
  \citenamefont {Goian}, \citenamefont {Rocas}, \citenamefont {Haislmaier},
  \citenamefont {Vlahos}, \citenamefont {Mundy}, \citenamefont {Kourkoutis},
  \citenamefont {Nie}, \citenamefont {Biegalski}, \citenamefont {Zhang},
  \citenamefont {Bernhagen}, \citenamefont {Benedek}, \citenamefont {Kim},
  \citenamefont {Brock}, \citenamefont {Uecker}, \citenamefont {Xi},
  \citenamefont {Gopalan}, \citenamefont {Nuzhnyy}, \citenamefont {Kamba},
  \citenamefont {Muller}, \citenamefont {Takeuchi}, \citenamefont {Booth},
  \citenamefont {Fennie},\ and\ \citenamefont {Schlom}}]{lee2013exploiting}%
  \BibitemOpen
  \bibfield  {author} {\bibinfo {author} {\bibfnamefont {C.-H.}\ \bibnamefont
  {Lee}}, \bibinfo {author} {\bibfnamefont {N.~D.}\ \bibnamefont {Orloff}},
  \bibinfo {author} {\bibfnamefont {T.}~\bibnamefont {Birol}}, \bibinfo
  {author} {\bibfnamefont {Y.}~\bibnamefont {Zhu}}, \bibinfo {author}
  {\bibfnamefont {V.}~\bibnamefont {Goian}}, \bibinfo {author} {\bibfnamefont
  {E.}~\bibnamefont {Rocas}}, \bibinfo {author} {\bibfnamefont
  {R.}~\bibnamefont {Haislmaier}}, \bibinfo {author} {\bibfnamefont
  {E.}~\bibnamefont {Vlahos}}, \bibinfo {author} {\bibfnamefont {J.~A.}\
  \bibnamefont {Mundy}}, \bibinfo {author} {\bibfnamefont {L.~F.}\ \bibnamefont
  {Kourkoutis}}, \bibinfo {author} {\bibfnamefont {Y.}~\bibnamefont {Nie}},
  \bibinfo {author} {\bibfnamefont {M.~D.}\ \bibnamefont {Biegalski}}, \bibinfo
  {author} {\bibfnamefont {J.}~\bibnamefont {Zhang}}, \bibinfo {author}
  {\bibfnamefont {M.}~\bibnamefont {Bernhagen}}, \bibinfo {author}
  {\bibfnamefont {N.~A.}\ \bibnamefont {Benedek}}, \bibinfo {author}
  {\bibfnamefont {Y.}~\bibnamefont {Kim}}, \bibinfo {author} {\bibfnamefont
  {J.~D.}\ \bibnamefont {Brock}}, \bibinfo {author} {\bibfnamefont
  {R.}~\bibnamefont {Uecker}}, \bibinfo {author} {\bibfnamefont {X.~X.}\
  \bibnamefont {Xi}}, \bibinfo {author} {\bibfnamefont {V.}~\bibnamefont
  {Gopalan}}, \bibinfo {author} {\bibfnamefont {D.}~\bibnamefont {Nuzhnyy}},
  \bibinfo {author} {\bibfnamefont {S.}~\bibnamefont {Kamba}}, \bibinfo
  {author} {\bibfnamefont {D.~A.}\ \bibnamefont {Muller}}, \bibinfo {author}
  {\bibfnamefont {I.}~\bibnamefont {Takeuchi}}, \bibinfo {author}
  {\bibfnamefont {J.~C.}\ \bibnamefont {Booth}}, \bibinfo {author}
  {\bibfnamefont {C.~J.}\ \bibnamefont {Fennie}},\ and\ \bibinfo {author}
  {\bibfnamefont {D.~G.}\ \bibnamefont {Schlom}},\ }\bibfield  {title}
  {\bibinfo {title} {Exploiting dimensionality and defect mitigation to create
  tunable microwave dielectrics},\ }\href {https://doi.org/10.1038/nature12582}
  {\bibfield  {journal} {\bibinfo  {journal} {Nature}\ }\textbf {\bibinfo
  {volume} {502}},\ \bibinfo {pages} {532–536} (\bibinfo {year}
  {2013})}\BibitemShut {NoStop}%
\bibitem [{\citenamefont {Sun}\ \emph {et~al.}(2023)\citenamefont {Sun},
  \citenamefont {Huo}, \citenamefont {Hu}, \citenamefont {Li}, \citenamefont
  {Liu}, \citenamefont {Han}, \citenamefont {Tang}, \citenamefont {Mao},
  \citenamefont {Yang}, \citenamefont {Wang}, \citenamefont {Cheng},
  \citenamefont {Yao}, \citenamefont {Zhang},\ and\ \citenamefont
  {Wang}}]{sun2023signatures}%
  \BibitemOpen
  \bibfield  {author} {\bibinfo {author} {\bibfnamefont {H.}~\bibnamefont
  {Sun}}, \bibinfo {author} {\bibfnamefont {M.}~\bibnamefont {Huo}}, \bibinfo
  {author} {\bibfnamefont {X.}~\bibnamefont {Hu}}, \bibinfo {author}
  {\bibfnamefont {J.}~\bibnamefont {Li}}, \bibinfo {author} {\bibfnamefont
  {Z.}~\bibnamefont {Liu}}, \bibinfo {author} {\bibfnamefont {Y.}~\bibnamefont
  {Han}}, \bibinfo {author} {\bibfnamefont {L.}~\bibnamefont {Tang}}, \bibinfo
  {author} {\bibfnamefont {Z.}~\bibnamefont {Mao}}, \bibinfo {author}
  {\bibfnamefont {P.}~\bibnamefont {Yang}}, \bibinfo {author} {\bibfnamefont
  {B.}~\bibnamefont {Wang}}, \bibinfo {author} {\bibfnamefont {J.}~\bibnamefont
  {Cheng}}, \bibinfo {author} {\bibfnamefont {D.-X.}\ \bibnamefont {Yao}},
  \bibinfo {author} {\bibfnamefont {G.-M.}\ \bibnamefont {Zhang}},\ and\
  \bibinfo {author} {\bibfnamefont {M.}~\bibnamefont {Wang}},\ }\bibfield
  {title} {\bibinfo {title} {Signatures of superconductivity near 80 k in a
  nickelate under high pressure},\ }\href
  {https://doi.org/10.1038/s41586-023-06408-7} {\bibfield  {journal} {\bibinfo
  {journal} {Nature}\ }\textbf {\bibinfo {volume} {621}},\ \bibinfo {pages}
  {493–498} (\bibinfo {year} {2023})}\BibitemShut {NoStop}%
\bibitem [{ko2()}]{ko2024signatures}%
  \BibitemOpen
  \bibfield  {title} {\bibinfo {title} {Signatures of ambient pressure
  superconductivity in thin film {$\mathrm{La}_3\mathrm{Ni}_2\mathrm{O}_7$},
  volume = {638}, issn = {1476-4687}, url =
  {http://dx.doi.org/10.1038/s41586-024-08525-3}, doi =
  {10.1038/s41586-024-08525-3}, number = {8052}, journal = {Nature}, publisher
  = {Springer Science and Business Media LLC}, author = {Ko, Eun Kyo and Yu,
  Yijun and Liu, Yidi and Bhatt, Lopa and Li, Jiarui and Thampy, Vivek and Kuo,
  Cheng-Tai and Wang, Bai Yang and Lee, Yonghun and Lee, Kyuho and Lee, Jun-Sik
  and Goodge, Berit H. and Muller, David A. and Hwang, Harold Y.}, year =
  {2024}, month = dec, pages = {935–940}},\ }\href@noop {} {\ }\BibitemShut
  {NoStop}%
\bibitem [{\citenamefont {Wang}\ \emph {et~al.}(2024)\citenamefont {Wang},
  \citenamefont {Wang}, \citenamefont {Shen}, \citenamefont {Hou},
  \citenamefont {Luo}, \citenamefont {Ma}, \citenamefont {Yang}, \citenamefont
  {Shi}, \citenamefont {Dou}, \citenamefont {Feng}, \citenamefont {Yang},
  \citenamefont {Shi}, \citenamefont {Ren}, \citenamefont {Ma}, \citenamefont
  {Yang}, \citenamefont {Liu}, \citenamefont {Liu}, \citenamefont {Zhang},
  \citenamefont {Dong}, \citenamefont {Wang}, \citenamefont {Jiang},
  \citenamefont {Hu}, \citenamefont {Nagasaki}, \citenamefont {Kitagawa},
  \citenamefont {Calder}, \citenamefont {Yan}, \citenamefont {Sun},
  \citenamefont {Wang}, \citenamefont {Zhou}, \citenamefont {Uwatoko},\ and\
  \citenamefont {Cheng}}]{wang2024bulk}%
  \BibitemOpen
  \bibfield  {author} {\bibinfo {author} {\bibfnamefont {N.}~\bibnamefont
  {Wang}}, \bibinfo {author} {\bibfnamefont {G.}~\bibnamefont {Wang}}, \bibinfo
  {author} {\bibfnamefont {X.}~\bibnamefont {Shen}}, \bibinfo {author}
  {\bibfnamefont {J.}~\bibnamefont {Hou}}, \bibinfo {author} {\bibfnamefont
  {J.}~\bibnamefont {Luo}}, \bibinfo {author} {\bibfnamefont {X.}~\bibnamefont
  {Ma}}, \bibinfo {author} {\bibfnamefont {H.}~\bibnamefont {Yang}}, \bibinfo
  {author} {\bibfnamefont {L.}~\bibnamefont {Shi}}, \bibinfo {author}
  {\bibfnamefont {J.}~\bibnamefont {Dou}}, \bibinfo {author} {\bibfnamefont
  {J.}~\bibnamefont {Feng}}, \bibinfo {author} {\bibfnamefont {J.}~\bibnamefont
  {Yang}}, \bibinfo {author} {\bibfnamefont {Y.}~\bibnamefont {Shi}}, \bibinfo
  {author} {\bibfnamefont {Z.}~\bibnamefont {Ren}}, \bibinfo {author}
  {\bibfnamefont {H.}~\bibnamefont {Ma}}, \bibinfo {author} {\bibfnamefont
  {P.}~\bibnamefont {Yang}}, \bibinfo {author} {\bibfnamefont {Z.}~\bibnamefont
  {Liu}}, \bibinfo {author} {\bibfnamefont {Y.}~\bibnamefont {Liu}}, \bibinfo
  {author} {\bibfnamefont {H.}~\bibnamefont {Zhang}}, \bibinfo {author}
  {\bibfnamefont {X.}~\bibnamefont {Dong}}, \bibinfo {author} {\bibfnamefont
  {Y.}~\bibnamefont {Wang}}, \bibinfo {author} {\bibfnamefont {K.}~\bibnamefont
  {Jiang}}, \bibinfo {author} {\bibfnamefont {J.}~\bibnamefont {Hu}}, \bibinfo
  {author} {\bibfnamefont {S.}~\bibnamefont {Nagasaki}}, \bibinfo {author}
  {\bibfnamefont {K.}~\bibnamefont {Kitagawa}}, \bibinfo {author}
  {\bibfnamefont {S.}~\bibnamefont {Calder}}, \bibinfo {author} {\bibfnamefont
  {J.}~\bibnamefont {Yan}}, \bibinfo {author} {\bibfnamefont {J.}~\bibnamefont
  {Sun}}, \bibinfo {author} {\bibfnamefont {B.}~\bibnamefont {Wang}}, \bibinfo
  {author} {\bibfnamefont {R.}~\bibnamefont {Zhou}}, \bibinfo {author}
  {\bibfnamefont {Y.}~\bibnamefont {Uwatoko}},\ and\ \bibinfo {author}
  {\bibfnamefont {J.}~\bibnamefont {Cheng}},\ }\bibfield  {title} {\bibinfo
  {title} {Bulk high-temperature superconductivity in pressurized tetragonal
  {$\mathrm{La}_2\mathrm{Pr}\mathrm{Ni}_2\mathrm{O}_7$}},\ }\href
  {https://doi.org/10.1038/s41586-024-07996-8} {\bibfield  {journal} {\bibinfo
  {journal} {Nature}\ }\textbf {\bibinfo {volume} {634}},\ \bibinfo {pages}
  {579–584} (\bibinfo {year} {2024})}\BibitemShut {NoStop}%
\bibitem [{\citenamefont {Zhou}\ \emph {et~al.}(2024)\citenamefont {Zhou},
  \citenamefont {Lv}, \citenamefont {Wang}, \citenamefont {Nie}, \citenamefont
  {Chen}, \citenamefont {Li}, \citenamefont {Huang}, \citenamefont {Chen},
  \citenamefont {Sun}, \citenamefont {Xue},\ and\ \citenamefont
  {Chen}}]{zhou2024ambient}%
  \BibitemOpen
  \bibfield  {author} {\bibinfo {author} {\bibfnamefont {G.}~\bibnamefont
  {Zhou}}, \bibinfo {author} {\bibfnamefont {W.}~\bibnamefont {Lv}}, \bibinfo
  {author} {\bibfnamefont {H.}~\bibnamefont {Wang}}, \bibinfo {author}
  {\bibfnamefont {Z.}~\bibnamefont {Nie}}, \bibinfo {author} {\bibfnamefont
  {Y.}~\bibnamefont {Chen}}, \bibinfo {author} {\bibfnamefont {Y.}~\bibnamefont
  {Li}}, \bibinfo {author} {\bibfnamefont {H.}~\bibnamefont {Huang}}, \bibinfo
  {author} {\bibfnamefont {W.}~\bibnamefont {Chen}}, \bibinfo {author}
  {\bibfnamefont {Y.}~\bibnamefont {Sun}}, \bibinfo {author} {\bibfnamefont
  {Q.-K.}\ \bibnamefont {Xue}},\ and\ \bibinfo {author} {\bibfnamefont
  {Z.}~\bibnamefont {Chen}},\ }\bibfield  {title} {\bibinfo {title}
  {Ambient-pressure superconductivity onset above 40 k in bilayer nickelate
  ultrathin films}\ }\href {https://doi.org/10.48550/arxiv.2412.16622}
  {10.48550/arxiv.2412.16622} (\bibinfo {year} {2024})\BibitemShut {NoStop}%
\bibitem [{\citenamefont {Zhou}\ \emph {et~al.}(2025)\citenamefont {Zhou},
  \citenamefont {Shu}, \citenamefont {Zhang}, \citenamefont {Liu},
  \citenamefont {Liu}, \citenamefont {Xiao}, \citenamefont {Shen},
  \citenamefont {Wu}, \citenamefont {Li}, \citenamefont {Zhang}, \citenamefont
  {Lyu}, \citenamefont {Wu}, \citenamefont {Sabri}, \citenamefont {Wang},
  \citenamefont {Yi}, \citenamefont {Nan}, \citenamefont {Zhang}, \citenamefont
  {He}, \citenamefont {Zang}, \citenamefont {Yang}, \citenamefont {Zhou},
  \citenamefont {Chen},\ and\ \citenamefont {Yu}}]{Yu}%
  \BibitemOpen
  \bibfield  {author} {\bibinfo {author} {\bibfnamefont {Y.}~\bibnamefont
  {Zhou}}, \bibinfo {author} {\bibfnamefont {X.}~\bibnamefont {Shu}}, \bibinfo
  {author} {\bibfnamefont {Y.}~\bibnamefont {Zhang}}, \bibinfo {author}
  {\bibfnamefont {Z.}~\bibnamefont {Liu}}, \bibinfo {author} {\bibfnamefont
  {L.}~\bibnamefont {Liu}}, \bibinfo {author} {\bibfnamefont {K.}~\bibnamefont
  {Xiao}}, \bibinfo {author} {\bibfnamefont {S.}~\bibnamefont {Shen}}, \bibinfo
  {author} {\bibfnamefont {S.}~\bibnamefont {Wu}}, \bibinfo {author}
  {\bibfnamefont {C.}~\bibnamefont {Li}}, \bibinfo {author} {\bibfnamefont
  {J.}~\bibnamefont {Zhang}}, \bibinfo {author} {\bibfnamefont
  {Y.}~\bibnamefont {Lyu}}, \bibinfo {author} {\bibfnamefont {Y.}~\bibnamefont
  {Wu}}, \bibinfo {author} {\bibfnamefont {H.}~\bibnamefont {Sabri}}, \bibinfo
  {author} {\bibfnamefont {M.}~\bibnamefont {Wang}}, \bibinfo {author}
  {\bibfnamefont {D.}~\bibnamefont {Yi}}, \bibinfo {author} {\bibfnamefont
  {T.}~\bibnamefont {Nan}}, \bibinfo {author} {\bibfnamefont {G.-M.}\
  \bibnamefont {Zhang}}, \bibinfo {author} {\bibfnamefont {Q.}~\bibnamefont
  {He}}, \bibinfo {author} {\bibfnamefont {J.}~\bibnamefont {Zang}}, \bibinfo
  {author} {\bibfnamefont {L.}~\bibnamefont {Yang}}, \bibinfo {author}
  {\bibfnamefont {S.}~\bibnamefont {Zhou}}, \bibinfo {author} {\bibfnamefont
  {H.}~\bibnamefont {Chen}},\ and\ \bibinfo {author} {\bibfnamefont
  {P.}~\bibnamefont {Yu}},\ }\bibfield  {title} {\bibinfo {title}
  {Geometry-driven polar antiferromagnetic metallicity in a double-layered
  perovskite cobaltate},\ }\bibfield  {journal} {\bibinfo  {journal} {Nature
  Materials}\ }\href {https://doi.org/10.1038/s41563-025-02392-7}
  {10.1038/s41563-025-02392-7} (\bibinfo {year} {2025})\BibitemShut {NoStop}%
\bibitem [{\citenamefont {Shen}\ \emph {et~al.}(2023)\citenamefont {Shen},
  \citenamefont {Qin},\ and\ \citenamefont {Zhang}}]{shen2023effective}%
  \BibitemOpen
  \bibfield  {author} {\bibinfo {author} {\bibfnamefont {Y.}~\bibnamefont
  {Shen}}, \bibinfo {author} {\bibfnamefont {M.}~\bibnamefont {Qin}},\ and\
  \bibinfo {author} {\bibfnamefont {G.-M.}\ \bibnamefont {Zhang}},\ }\bibfield
  {title} {\bibinfo {title} {Effective bi-layer model hamiltonian and
  density-matrix renormalization group study for the high-t c superconductivity
  in {$\mathrm{La}_3\mathrm{Ni}_2\mathrm{O}_7$} under high pressure},\
  }\href@noop {} {\bibfield  {journal} {\bibinfo  {journal} {Chinese Physics
  Letters}\ }\textbf {\bibinfo {volume} {40}},\ \bibinfo {pages} {127401}
  (\bibinfo {year} {2023})}\BibitemShut {NoStop}%
\bibitem [{\citenamefont {Yang}\ \emph {et~al.}(2023)\citenamefont {Yang},
  \citenamefont {Zhang},\ and\ \citenamefont {Zhang}}]{PhysRevB.108.L201108}%
  \BibitemOpen
  \bibfield  {author} {\bibinfo {author} {\bibfnamefont {Y.-f.}\ \bibnamefont
  {Yang}}, \bibinfo {author} {\bibfnamefont {G.-M.}\ \bibnamefont {Zhang}},\
  and\ \bibinfo {author} {\bibfnamefont {F.-C.}\ \bibnamefont {Zhang}},\
  }\bibfield  {title} {\bibinfo {title} {Interlayer valence bonds and
  two-component theory for high-${T}_{c}$ superconductivity of
  {${\mathrm{La}}_{3}{\mathrm{Ni}}_{2}{\mathrm{O}}_{7}$} under pressure},\
  }\href {https://doi.org/10.1103/PhysRevB.108.L201108} {\bibfield  {journal}
  {\bibinfo  {journal} {Phys. Rev. B}\ }\textbf {\bibinfo {volume} {108}},\
  \bibinfo {pages} {L201108} (\bibinfo {year} {2023})}\BibitemShut {NoStop}%
\bibitem [{\citenamefont {Sokolov}\ \emph {et~al.}(2019)\citenamefont
  {Sokolov}, \citenamefont {Kikugawa}, \citenamefont {Helm}, \citenamefont
  {Borrmann}, \citenamefont {Burkhardt}, \citenamefont {Cubitt}, \citenamefont
  {White}, \citenamefont {Ressouche}, \citenamefont {Bleuel}, \citenamefont
  {Kummer}, \citenamefont {Mackenzie},\ and\ \citenamefont
  {R\"{o}ßler}}]{sokolov2019metamagnetic}%
  \BibitemOpen
  \bibfield  {author} {\bibinfo {author} {\bibfnamefont {D.~A.}\ \bibnamefont
  {Sokolov}}, \bibinfo {author} {\bibfnamefont {N.}~\bibnamefont {Kikugawa}},
  \bibinfo {author} {\bibfnamefont {T.}~\bibnamefont {Helm}}, \bibinfo {author}
  {\bibfnamefont {H.}~\bibnamefont {Borrmann}}, \bibinfo {author}
  {\bibfnamefont {U.}~\bibnamefont {Burkhardt}}, \bibinfo {author}
  {\bibfnamefont {R.}~\bibnamefont {Cubitt}}, \bibinfo {author} {\bibfnamefont
  {J.~S.}\ \bibnamefont {White}}, \bibinfo {author} {\bibfnamefont
  {E.}~\bibnamefont {Ressouche}}, \bibinfo {author} {\bibfnamefont
  {M.}~\bibnamefont {Bleuel}}, \bibinfo {author} {\bibfnamefont
  {K.}~\bibnamefont {Kummer}}, \bibinfo {author} {\bibfnamefont {A.~P.}\
  \bibnamefont {Mackenzie}},\ and\ \bibinfo {author} {\bibfnamefont {U.~K.}\
  \bibnamefont {R\"{o}ßler}},\ }\bibfield  {title} {\bibinfo {title}
  {Metamagnetic texture in a polar antiferromagnet},\ }\href
  {https://doi.org/10.1038/s41567-019-0501-0} {\bibfield  {journal} {\bibinfo
  {journal} {Nature Physics}\ }\textbf {\bibinfo {volume} {15}},\ \bibinfo
  {pages} {671–677} (\bibinfo {year} {2019})}\BibitemShut {NoStop}%
\bibitem [{\citenamefont {Marković}\ \emph {et~al.}(2020)\citenamefont
  {Marković}, \citenamefont {Watson}, \citenamefont {Clark}, \citenamefont
  {Mazzola}, \citenamefont {Morales}, \citenamefont {Hooley}, \citenamefont
  {Rosner}, \citenamefont {Polley}, \citenamefont {Balasubramanian},
  \citenamefont {Mukherjee}, \citenamefont {Kikugawa}, \citenamefont {Sokolov},
  \citenamefont {Mackenzie},\ and\ \citenamefont
  {King}}]{doi:10.1073/pnas.2003671117}%
  \BibitemOpen
  \bibfield  {author} {\bibinfo {author} {\bibfnamefont {I.}~\bibnamefont
  {Marković}}, \bibinfo {author} {\bibfnamefont {M.~D.}\ \bibnamefont
  {Watson}}, \bibinfo {author} {\bibfnamefont {O.~J.}\ \bibnamefont {Clark}},
  \bibinfo {author} {\bibfnamefont {F.}~\bibnamefont {Mazzola}}, \bibinfo
  {author} {\bibfnamefont {E.~A.}\ \bibnamefont {Morales}}, \bibinfo {author}
  {\bibfnamefont {C.~A.}\ \bibnamefont {Hooley}}, \bibinfo {author}
  {\bibfnamefont {H.}~\bibnamefont {Rosner}}, \bibinfo {author} {\bibfnamefont
  {C.~M.}\ \bibnamefont {Polley}}, \bibinfo {author} {\bibfnamefont
  {T.}~\bibnamefont {Balasubramanian}}, \bibinfo {author} {\bibfnamefont
  {S.}~\bibnamefont {Mukherjee}}, \bibinfo {author} {\bibfnamefont
  {N.}~\bibnamefont {Kikugawa}}, \bibinfo {author} {\bibfnamefont {D.~A.}\
  \bibnamefont {Sokolov}}, \bibinfo {author} {\bibfnamefont {A.~P.}\
  \bibnamefont {Mackenzie}},\ and\ \bibinfo {author} {\bibfnamefont {P.~D.~C.}\
  \bibnamefont {King}},\ }\bibfield  {title} {\bibinfo {title} {Electronically
  driven spin-reorientation transition of the correlated polar metal
  $\mathrm{Ca}_3\mathrm{Ru}_2\mathrm{O}_7$},\ }\href
  {https://doi.org/10.1073/pnas.2003671117} {\bibfield  {journal} {\bibinfo
  {journal} {Proceedings of the National Academy of Sciences}\ }\textbf
  {\bibinfo {volume} {117}},\ \bibinfo {pages} {15524} (\bibinfo {year}
  {2020})},\ \Eprint
  {https://arxiv.org/abs/https://www.pnas.org/doi/pdf/10.1073/pnas.2003671117}
  {https://www.pnas.org/doi/pdf/10.1073/pnas.2003671117} \BibitemShut {NoStop}%
\bibitem [{\citenamefont {Peng}\ \emph {et~al.}(2024)\citenamefont {Peng},
  \citenamefont {Park}, \citenamefont {Roh}, \citenamefont {Mun}, \citenamefont
  {Ju}, \citenamefont {Kim}, \citenamefont {Ko}, \citenamefont {Liang},
  \citenamefont {Hahn}, \citenamefont {Zhang}, \citenamefont {Sanchez},
  \citenamefont {Walker}, \citenamefont {Hindmarsh}, \citenamefont {Si},
  \citenamefont {Jo}, \citenamefont {Jo}, \citenamefont {Kim}, \citenamefont
  {Kim}, \citenamefont {Wang}, \citenamefont {Kim}, \citenamefont {Lee},
  \citenamefont {Noh},\ and\ \citenamefont {Lee}}]{peng2024flexoelectric}%
  \BibitemOpen
  \bibfield  {author} {\bibinfo {author} {\bibfnamefont {W.}~\bibnamefont
  {Peng}}, \bibinfo {author} {\bibfnamefont {S.~Y.}\ \bibnamefont {Park}},
  \bibinfo {author} {\bibfnamefont {C.~J.}\ \bibnamefont {Roh}}, \bibinfo
  {author} {\bibfnamefont {J.}~\bibnamefont {Mun}}, \bibinfo {author}
  {\bibfnamefont {H.}~\bibnamefont {Ju}}, \bibinfo {author} {\bibfnamefont
  {J.}~\bibnamefont {Kim}}, \bibinfo {author} {\bibfnamefont {E.~K.}\
  \bibnamefont {Ko}}, \bibinfo {author} {\bibfnamefont {Z.}~\bibnamefont
  {Liang}}, \bibinfo {author} {\bibfnamefont {S.}~\bibnamefont {Hahn}},
  \bibinfo {author} {\bibfnamefont {J.}~\bibnamefont {Zhang}}, \bibinfo
  {author} {\bibfnamefont {A.~M.}\ \bibnamefont {Sanchez}}, \bibinfo {author}
  {\bibfnamefont {D.}~\bibnamefont {Walker}}, \bibinfo {author} {\bibfnamefont
  {S.}~\bibnamefont {Hindmarsh}}, \bibinfo {author} {\bibfnamefont
  {L.}~\bibnamefont {Si}}, \bibinfo {author} {\bibfnamefont {Y.~J.}\
  \bibnamefont {Jo}}, \bibinfo {author} {\bibfnamefont {Y.}~\bibnamefont {Jo}},
  \bibinfo {author} {\bibfnamefont {T.~H.}\ \bibnamefont {Kim}}, \bibinfo
  {author} {\bibfnamefont {C.}~\bibnamefont {Kim}}, \bibinfo {author}
  {\bibfnamefont {L.}~\bibnamefont {Wang}}, \bibinfo {author} {\bibfnamefont
  {M.}~\bibnamefont {Kim}}, \bibinfo {author} {\bibfnamefont {J.~S.}\
  \bibnamefont {Lee}}, \bibinfo {author} {\bibfnamefont {T.~W.}\ \bibnamefont
  {Noh}},\ and\ \bibinfo {author} {\bibfnamefont {D.}~\bibnamefont {Lee}},\
  }\bibfield  {title} {\bibinfo {title} {Flexoelectric polarizing and control
  of a ferromagnetic metal},\ }\href
  {https://doi.org/10.1038/s41567-023-02333-8} {\bibfield  {journal} {\bibinfo
  {journal} {Nature Physics}\ }\textbf {\bibinfo {volume} {20}},\ \bibinfo
  {pages} {450–455} (\bibinfo {year} {2024})}\BibitemShut {NoStop}%
\bibitem [{\citenamefont {Gui}\ \emph {et~al.}(2022)\citenamefont {Gui},
  \citenamefont {Gu}, \citenamefont {Cheng}, \citenamefont {Zhu}, \citenamefont
  {Yu}, \citenamefont {Guo}, \citenamefont {Wu}, \citenamefont {Mei},
  \citenamefont {Sheng}, \citenamefont {Zhang}, \citenamefont {Wang},
  \citenamefont {Zhao}, \citenamefont {Bellaiche}, \citenamefont {Huang},\ and\
  \citenamefont {Wang}}]{gui2022improper}%
  \BibitemOpen
  \bibfield  {author} {\bibinfo {author} {\bibfnamefont {Z.}~\bibnamefont
  {Gui}}, \bibinfo {author} {\bibfnamefont {C.}~\bibnamefont {Gu}}, \bibinfo
  {author} {\bibfnamefont {H.}~\bibnamefont {Cheng}}, \bibinfo {author}
  {\bibfnamefont {J.}~\bibnamefont {Zhu}}, \bibinfo {author} {\bibfnamefont
  {X.}~\bibnamefont {Yu}}, \bibinfo {author} {\bibfnamefont {E.-j.}\
  \bibnamefont {Guo}}, \bibinfo {author} {\bibfnamefont {L.}~\bibnamefont
  {Wu}}, \bibinfo {author} {\bibfnamefont {J.}~\bibnamefont {Mei}}, \bibinfo
  {author} {\bibfnamefont {J.}~\bibnamefont {Sheng}}, \bibinfo {author}
  {\bibfnamefont {J.}~\bibnamefont {Zhang}}, \bibinfo {author} {\bibfnamefont
  {J.}~\bibnamefont {Wang}}, \bibinfo {author} {\bibfnamefont {Y.}~\bibnamefont
  {Zhao}}, \bibinfo {author} {\bibfnamefont {L.}~\bibnamefont {Bellaiche}},
  \bibinfo {author} {\bibfnamefont {L.}~\bibnamefont {Huang}},\ and\ \bibinfo
  {author} {\bibfnamefont {S.}~\bibnamefont {Wang}},\ }\bibfield  {title}
  {\bibinfo {title} {Improper multiferroiclike transition in a metal},\ }\href
  {https://doi.org/10.1103/PhysRevB.105.L180101} {\bibfield  {journal}
  {\bibinfo  {journal} {Phys. Rev. B}\ }\textbf {\bibinfo {volume} {105}},\
  \bibinfo {pages} {L180101} (\bibinfo {year} {2022})}\BibitemShut {NoStop}%
\bibitem [{\citenamefont {Khomskii}(2014)}]{khomskii2014transition}%
  \BibitemOpen
  \bibfield  {author} {\bibinfo {author} {\bibfnamefont {D.~I.}\ \bibnamefont
  {Khomskii}},\ }\href@noop {} {\emph {\bibinfo {title} {Transition metal
  compounds}}}\ (\bibinfo  {publisher} {Cambridge University Press},\ \bibinfo
  {year} {2014})\BibitemShut {NoStop}%
\bibitem [{\citenamefont {Zaanen}\ \emph {et~al.}(1985)\citenamefont {Zaanen},
  \citenamefont {Sawatzky},\ and\ \citenamefont {Allen}}]{zaanen1985band}%
  \BibitemOpen
  \bibfield  {author} {\bibinfo {author} {\bibfnamefont {J.}~\bibnamefont
  {Zaanen}}, \bibinfo {author} {\bibfnamefont {G.}~\bibnamefont {Sawatzky}},\
  and\ \bibinfo {author} {\bibfnamefont {J.}~\bibnamefont {Allen}},\ }\bibfield
   {title} {\bibinfo {title} {Band gaps and electronic structure of
  transition-metal compounds},\ }\href@noop {} {\bibfield  {journal} {\bibinfo
  {journal} {Physical review letters}\ }\textbf {\bibinfo {volume} {55}},\
  \bibinfo {pages} {418} (\bibinfo {year} {1985})}\BibitemShut {NoStop}%
\bibitem [{\citenamefont {Wang}\ \emph {et~al.}(2020)\citenamefont {Wang},
  \citenamefont {He}, \citenamefont {Ming}, \citenamefont {Du}, \citenamefont
  {Lu}, \citenamefont {Cafolla}, \citenamefont {Fujioka}, \citenamefont
  {Zhang}, \citenamefont {Zhang}, \citenamefont {Shen}, \citenamefont {Lyu},
  \citenamefont {N'Diaye}, \citenamefont {Arenholz}, \citenamefont {Gu},
  \citenamefont {Nan}, \citenamefont {Tokura}, \citenamefont {Okamoto},\ and\
  \citenamefont {Yu}}]{PhysRevX.10.021030}%
  \BibitemOpen
  \bibfield  {author} {\bibinfo {author} {\bibfnamefont {Y.}~\bibnamefont
  {Wang}}, \bibinfo {author} {\bibfnamefont {Q.}~\bibnamefont {He}}, \bibinfo
  {author} {\bibfnamefont {W.}~\bibnamefont {Ming}}, \bibinfo {author}
  {\bibfnamefont {M.-H.}\ \bibnamefont {Du}}, \bibinfo {author} {\bibfnamefont
  {N.}~\bibnamefont {Lu}}, \bibinfo {author} {\bibfnamefont {C.}~\bibnamefont
  {Cafolla}}, \bibinfo {author} {\bibfnamefont {J.}~\bibnamefont {Fujioka}},
  \bibinfo {author} {\bibfnamefont {Q.}~\bibnamefont {Zhang}}, \bibinfo
  {author} {\bibfnamefont {D.}~\bibnamefont {Zhang}}, \bibinfo {author}
  {\bibfnamefont {S.}~\bibnamefont {Shen}}, \bibinfo {author} {\bibfnamefont
  {Y.}~\bibnamefont {Lyu}}, \bibinfo {author} {\bibfnamefont {A.~T.}\
  \bibnamefont {N'Diaye}}, \bibinfo {author} {\bibfnamefont {E.}~\bibnamefont
  {Arenholz}}, \bibinfo {author} {\bibfnamefont {L.}~\bibnamefont {Gu}},
  \bibinfo {author} {\bibfnamefont {C.}~\bibnamefont {Nan}}, \bibinfo {author}
  {\bibfnamefont {Y.}~\bibnamefont {Tokura}}, \bibinfo {author} {\bibfnamefont
  {S.}~\bibnamefont {Okamoto}},\ and\ \bibinfo {author} {\bibfnamefont
  {P.}~\bibnamefont {Yu}},\ }\bibfield  {title} {\bibinfo {title} {Robust
  ferromagnetism in highly strained {$\mathrm{Sr}\mathrm{Co}\mathrm{O}_{3}$}
  thin films},\ }\href {https://doi.org/10.1103/PhysRevX.10.021030} {\bibfield
  {journal} {\bibinfo  {journal} {Phys. Rev. X}\ }\textbf {\bibinfo {volume}
  {10}},\ \bibinfo {pages} {021030} (\bibinfo {year} {2020})}\BibitemShut
  {NoStop}%
\bibitem [{\citenamefont {Johnston}\ \emph {et~al.}(2014)\citenamefont
  {Johnston}, \citenamefont {Mukherjee}, \citenamefont {Elfimov}, \citenamefont
  {Berciu},\ and\ \citenamefont {Sawatzky}}]{PhysRevLett.112.106404}%
  \BibitemOpen
  \bibfield  {author} {\bibinfo {author} {\bibfnamefont {S.}~\bibnamefont
  {Johnston}}, \bibinfo {author} {\bibfnamefont {A.}~\bibnamefont {Mukherjee}},
  \bibinfo {author} {\bibfnamefont {I.}~\bibnamefont {Elfimov}}, \bibinfo
  {author} {\bibfnamefont {M.}~\bibnamefont {Berciu}},\ and\ \bibinfo {author}
  {\bibfnamefont {G.~A.}\ \bibnamefont {Sawatzky}},\ }\bibfield  {title}
  {\bibinfo {title} {Charge disproportionation without charge transfer in the
  rare-earth-element nickelates as a possible mechanism for the metal-insulator
  transition},\ }\href {https://doi.org/10.1103/PhysRevLett.112.106404}
  {\bibfield  {journal} {\bibinfo  {journal} {Phys. Rev. Lett.}\ }\textbf
  {\bibinfo {volume} {112}},\ \bibinfo {pages} {106404} (\bibinfo {year}
  {2014})}\BibitemShut {NoStop}%
\bibitem [{\citenamefont {Zhang}\ \emph {et~al.}(2020)\citenamefont {Zhang},
  \citenamefont {Yang},\ and\ \citenamefont {Zhang}}]{zhang2020self}%
  \BibitemOpen
  \bibfield  {author} {\bibinfo {author} {\bibfnamefont {G.-M.}\ \bibnamefont
  {Zhang}}, \bibinfo {author} {\bibfnamefont {Y.-f.}\ \bibnamefont {Yang}},\
  and\ \bibinfo {author} {\bibfnamefont {F.-C.}\ \bibnamefont {Zhang}},\
  }\bibfield  {title} {\bibinfo {title} {Self-doped mott insulator for parent
  compounds of nickelate superconductors},\ }\href@noop {} {\bibfield
  {journal} {\bibinfo  {journal} {Physical Review B}\ }\textbf {\bibinfo
  {volume} {101}},\ \bibinfo {pages} {020501} (\bibinfo {year}
  {2020})}\BibitemShut {NoStop}%
\bibitem [{\citenamefont {Nomura}\ and\ \citenamefont
  {Arita}(2022)}]{nomura2022superconductivity}%
  \BibitemOpen
  \bibfield  {author} {\bibinfo {author} {\bibfnamefont {Y.}~\bibnamefont
  {Nomura}}\ and\ \bibinfo {author} {\bibfnamefont {R.}~\bibnamefont {Arita}},\
  }\bibfield  {title} {\bibinfo {title} {Superconductivity in infinite-layer
  nickelates},\ }\href@noop {} {\bibfield  {journal} {\bibinfo  {journal}
  {Reports on Progress in Physics}\ }\textbf {\bibinfo {volume} {85}},\
  \bibinfo {pages} {052501} (\bibinfo {year} {2022})}\BibitemShut {NoStop}%
\bibitem [{\citenamefont {Liechtenstein}\ \emph {et~al.}(1995)\citenamefont
  {Liechtenstein}, \citenamefont {Anisimov},\ and\ \citenamefont
  {Zaanen}}]{PhysRevB.52.R5467}%
  \BibitemOpen
  \bibfield  {author} {\bibinfo {author} {\bibfnamefont {A.~I.}\ \bibnamefont
  {Liechtenstein}}, \bibinfo {author} {\bibfnamefont {V.~I.}\ \bibnamefont
  {Anisimov}},\ and\ \bibinfo {author} {\bibfnamefont {J.}~\bibnamefont
  {Zaanen}},\ }\bibfield  {title} {\bibinfo {title} {Density-functional theory
  and strong interactions: Orbital ordering in mott-hubbard insulators},\
  }\href {https://doi.org/10.1103/PhysRevB.52.R5467} {\bibfield  {journal}
  {\bibinfo  {journal} {Phys. Rev. B}\ }\textbf {\bibinfo {volume} {52}},\
  \bibinfo {pages} {R5467} (\bibinfo {year} {1995})}\BibitemShut {NoStop}%
\end{thebibliography}%

\end{document}